\documentclass[journal]{IEEEtran}

\usepackage{lineno}
\usepackage{subfigure}
\usepackage{epsfig,graphics,subfigure,psfrag,amsmath,amssymb}
\usepackage{amsmath,amssymb,amsfonts}
\usepackage{algorithmic}
\usepackage{graphicx}
\usepackage{textcomp}
\usepackage[vlined,boxed,ruled]{algorithm2e}
\usepackage{dsfont}
\newtheorem{definition}{Definition}
\newtheorem{theorem}{Theorem}

\ifCLASSINFOpdf
\else

\fi

\begin{document}

\title{PRIF: A Privacy-Preserving Interest-Based Forwarding Scheme for Social Internet of Vehicles}

\author{Liehuang~Zhu,~\IEEEmembership{Member,~IEEE,}
        Chuan~Zhang, ~Chang~Xu, ~Xiaojiang~Du,~\IEEEmembership{Senior Member,~IEEE}, ~Rixin~Xu, ~Kashif~Sharif,~\IEEEmembership{Member,~IEEE}, and Mohsen Guizani,~\IEEEmembership{Fellow,~IEEE}
        \thanks{This research is supported by the National Natural Science
        	Foundation of China (Grant Nos. 61402037, 61272512). {\em(Corresponding author: Chang Xu.)}
        }
      \thanks{Liehuang Zhu, Chuan Zhang, Chang Xu, Rixin Xu, and Kashif~Sharif are with the Beijing Engineering Research Center of Massive Language Information Processing and Cloud Computing Application, School of Computer Science and Technology, Beijing Institute of Technology, Beijing, China, email: \{liehuangz, chuanz, xuchang, xurixin, 7620160009\}@bit.edu.cn.}
\thanks{Xiaojiang Du is with the Department of Computer and Information Sciences, Temple University, Philadelphia, USA, email: dxj2005@gmail.com.}
\thanks{Mohsen Guizani is with the Department of Electrical and Computer Engineering, University of Idaho, Moscow, Idaho, USA, e-mail: mguizani@ieee.org}
}

%
%

\markboth{IEEE INTERNET OF THINGS JOURNAL}%
{Shell \MakeLowercase{\textit{et al.}}: Bare Demo of IEEEtran.cls for IEEE Journals}
%



\maketitle

\begin{abstract}
Recent advances in Socially Aware Networks (SANs) have allowed its use in many domains, out of which social Internet of vehicles (SIOV) is of prime importance. SANs can provide a promising routing and forwarding paradigm for SIOV by using interest-based communication. Though able to improve the forwarding performance, existing interest-based schemes fail to consider the important issue of protecting users' interest information. In this paper, we propose a PRivacy-preserving Interest-based Forwarding scheme (PRIF) for SIOV, which not only protects the interest information, but also improves the forwarding performance. We propose a privacy-preserving authentication protocol to recognize communities among mobile nodes. During data routing and forwarding, a node can know others' interests only if they are affiliated with the same community. Moreover, to improve forwarding performance, a new metric {\em community energy} is introduced to indicate vehicular social proximity. Community energy is generated when two nodes encounter one another and information is shared among them. PRIF considers this energy metric to select forwarders towards the destination node or the destination community. Security analysis indicates PRIF can protect nodes' interest information. In addition, extensive simulations have been conducted to demonstrate that PRIF outperforms the existing algorithms including the BEEINFO, Epidemic,  and PRoPHET.
\end{abstract}

\begin{IEEEkeywords}
interest, privacy-preserving, forwarding, community, social Internet of vehicles.
\end{IEEEkeywords}

%
\IEEEpeerreviewmaketitle

\section{Introduction}
\IEEEPARstart{T}{he} recent proliferation of mobile devices (e.g., mobile phones, vehicle onboard equipment, tablets, etc.) has changed the future of communication and services \cite{Yuan2013}. Due to the inseparable bond between mobile devices and their human carriers, social relationships and users' mobility aspects are exploited in various research fields, such as Opportunistic Networks (OppNets) \cite{pelusi2006opportunistic}, Vehicular Ad-hoc Networks \cite{Lin2015}, and Delay Tolerant Networks \cite{Warthman2003}. This emerging network paradigm, also known as Socially Aware Networking (SAN), is able to take advantage of users' social properties, and further uses them as a main design ingredient for social Internet of vehicles (SIOV), a communication network where vehicles behave ``socially" \cite{vegni2015survey}, \cite{Hu2015S}.

SIOV works similarly to OppNets and DTNs, as they all lack end-to-end fixed paths from the source to the destination, they utilize store-carry-forward paradigm for such services. When there is a need for data dissemination, a key problem for these networks is to predict the future encounter opportunity. Nevertheless, the difference between SIOV and OppNets (or DTNs) is that SIOV considers social properties of devices to solve the data routing and forwarding problems and challenges.

\begin{figure}
	\centering
	\includegraphics[ width=10cm]{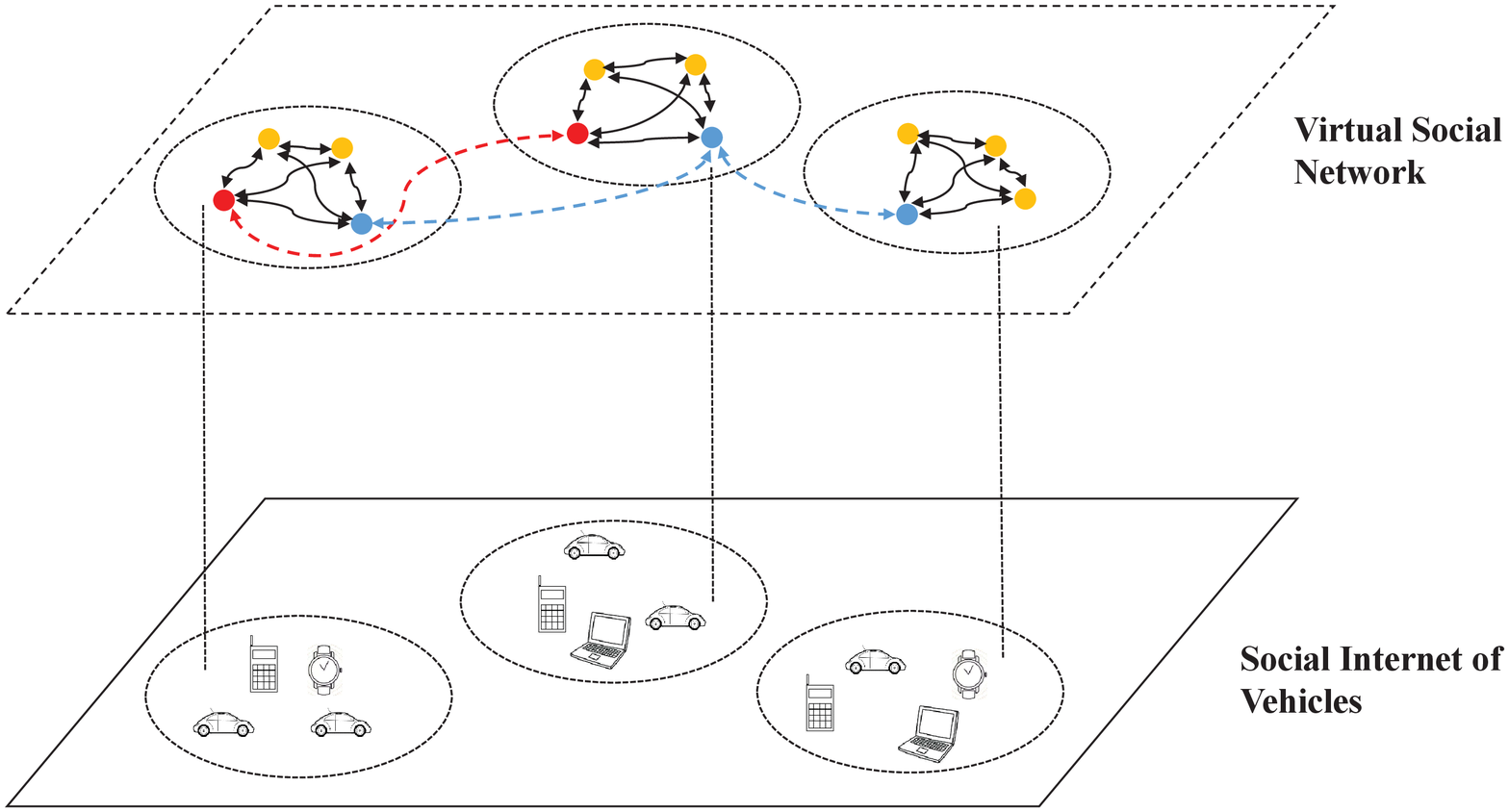}
	\caption{Virtual Social Networks and Social Internet of Vehicles. Mobile devices (red, blue and yellow circles) form electronic social networks when they are in proximity, while vehicles (red and blue circles) establish social ties based on contact frequency or common interests.}
	\label{fv}
\end{figure}

Fig. \ref{fv} depicts two social levels in vehicular environment. Vehicles/individuals carrying mobile devices (e.g., smartphone, smart watch, digital camera, etc.) construct social Internet of vehicles when they are in communication range, and their inherent social ties determine virtual social networks. Generally, social relationships are relatively stable and change less frequently than transmission links among mobile devices. Therefore, it is crucial to take advantage of mobile devices' social properties to make smarter forwarding decisions.

Recently, a series of social-based routing protocols, \cite{li2009localcom}, \cite{musolesi2008writing}, \cite{musolesi2005adaptive}, \cite{hui2007small}, \cite{hui2011bubble}, \cite{daly2007social}, \cite{bulut2010friendship}, \cite{li2014routing}, \cite{xia2015beeinfo}, \cite{daly2009social}, \cite{boldrini2008exploiting} have been proposed. Most of them adopt the notion of ``community" to make forwarding decisions. Specifically, mobile nodes can be divided into different communities based on their contact frequency or social relationships. It is generally agreed that members in a same community meet each other more often than others in different communities. Thus, a forwarding decision usually relies on how to construct a community and choose suitable forwarders. For example,  LocalCom \cite{li2009localcom} detects communities using neighboring graphs, while Gently \cite{musolesi2008writing} chooses forwarders based on CAR-like \cite{musolesi2005adaptive} and Label \cite{hui2007small} protocols. Communities in these schemes can be obtained from historical records such as encounter frequency, encounter length, and separation time. However, they ignore nodes' inherent social relationships, especially considering that mobile nodes are always carried and used by people. Many forwarding schemes \cite{hui2007small}, \cite{hui2011bubble}, \cite{daly2007social}, \cite{bulut2010friendship}, \cite{li2014routing} have been proposed based on social network metrics. For example, Label \cite{bulut2010friendship} and Group \cite{li2014routing} deliver messages only if message carriers meet the members within the same social community of the destination node, while Bubble Rap \cite{hui2011bubble} uses a hierarchical community structure and forwards data if a node holds higher centrality. These social-based works utilize social relationships to make better forwarding strategies. Nevertheless, the drawback in these schemes is that the cost to form and maintain communities is high. Recently, Xia et al. \cite{xia2015beeinfo} proposed a forwarding scheme named BEEINFO-D$\&$S which uses personal interests to construct communities. By using interest information, it eliminates the cost of community detection and formation. However, broadcasting the interest information is dangerous, since the interest usually contains users' sensitive information, which can be used to directly or indirectly determine trajectories, habits, and religious beliefs, etc.

In this paper, we propose PRIF: a PRivacy-preserving Interest-based Forwarding scheme to protect the sensitive interest information and improve the forwarding efficiency for SIOV. To summarize, the contributions of PRIF include:
\begin{itemize}
	\item First, we classify communities based on personal interests. Inspired by BEEINFO-D$\&$S and general laws in practical physics, a novel social metric {\em community energy} is introduced to measure the social ability of a mobile node to forward messages to others. Generally, community energy is generated by node encounters. Specifically, a node will establish inter-community energy to the encountering node, if it is within the same community. Otherwise, intra-community energy will be built towards the encountering node community. Therefore, a better forwarder should be a node with higher inter-community energy to the destination node, or a node with higher intra-community energy towards the destination community.
	
	\item Second, the interest information is private, and it is dangerous to deliver it to others. Thus, we take advantage of signature-based envelops and design a privacy-preserving authentication protocol. In this way, a node can  recognize the members coming from which communities. However,  it cannot   know the interests of the members unless they  are affiliated to the same group with it.
	
	\item Third, extensive simulation analysis has been conducted to compare PRIF with several existing schemes. Specifically, compared with two representative schemes, i.e., Epidemic \cite{epidemic} and PRoPHET \cite{PRoPHET}, PRIF performs better in message delivery, overhead, and hop counts. In addition, the proposed scheme outperforms the existing interest-based scheme BEEINFO-D$\&$S in delivery ratio and overhead.
\end{itemize}

The rest of this paper is organized as follows: In section II, we review existing social-based data forwarding protocols. Section III describes the detailed design of PRIF for SIOV. In section IV, we give the security analysis of PRIF and in section V, we analyze the performance. Finally, we conclude our work in section VI.

\section{Related works}
In recent years, feature extraction has received considerable attention in various fields \cite{thing2006}, \cite{He2016Connected}, \cite{He2017Robust},\cite{Du2017A}, \cite{Du2016Social}. In VANETs, to adapt to the frequently changing topology and high-speed mobility \cite{Du2015improving}, \cite{Du2016Energy}, \cite{DuC08}, social property as a special feature among people, plays an important role in designing routing algorithms. Many social-based routing algorithms \cite{li2009localcom}, \cite{musolesi2008writing}, \cite{musolesi2005adaptive}, \cite{hui2007small}, \cite{hui2011bubble}, \cite{daly2007social}, \cite{bulut2010friendship}, \cite{li2014routing}, \cite{xia2015beeinfo}, \cite{daly2009social}, \cite{boldrini2008exploiting} have been proposed which are roughly based on two main aspects: behavior regularity and community information.

Behavior regularity focuses on individuals' behaviors. It relies on the principle that people usually have repeated mobility patterns. In real world, people often hold similar mobility patterns. For instance, they usually follow   similar paths  from their home to offices during weekdays. Accordingly, regular behaviors can be used to predict the future encounter probability, and works in \cite{daly2007social}, \cite{daly2009social}, \cite{boldrini2008exploiting} have proposed   algorithms based on this metric. For example, SimBet \cite{daly2007social} constructs a utility function by exploiting the similarity and betweenness centrality to the destination with the help of an ego network. To describe nodes' relationships, SimBetTS \cite{daly2009social} considers another important factor (i.e., social tie strength) to choose more suitable forwarders. Moreover, HiBOp \cite{boldrini2008exploiting} can automatically learn users' behaviors and social relations to execute the forwarding process.

Another important basis to support social-based routing is community information. Generally, communities can be constructed by individuals' interests or encounter frequency, and it is generally agreed that nodes coming from the same community will meet each other more frequently. A series of routing algorithms have been proposed based on this metric \cite{li2009localcom}, \cite{musolesi2008writing}, \cite{hui2007small}, \cite{hui2011bubble},  \cite{bulut2010friendship}, \cite{li2014routing}, \cite{xia2015beeinfo}. The simplest community-based routing method is LABEL \cite{hui2007small}, in which messages are only delivered to the nodes in the destination community. Similar to the scheme in \cite{hui2007small}, Li et al. \cite{li2014routing} proposed a new community-based scheme. However, \cite{hui2007small} and \cite{li2014routing} do not take nodes' relationships into consideration. To solve the problem, Bubble Rap \cite{hui2011bubble} and Friendship-based \cite{bulut2010friendship} were proposed to select nodes with higher social centrality as relay nodes. Nevertheless, these schemes suffer a common drawback that the cost to form and maintain a socially aware overlay is extremely high. Besides the intra-community routing, inter-community routings were considered  in \cite{li2009localcom} and \cite{musolesi2008writing}. LocalCom \cite{li2009localcom} utilized the encounter history such as encounter frequency, encounter period and separation time to construct a neighboring graph which was further utilized to detect communities, represent nodes' similarity for intra-community, and design routing strategies for inter-community communication, while Gently \cite{musolesi2008writing} adopted a context-aware adaptive routing (CAR) algorithm and LABEL protocol. When no nodes within the destination community are in reach, Gently adopts the CAR-like routing algorithm. When the message carrier encounters a node coming from the destination community, it utilizes a LABEL-based protocol. Finally, a CAR-like routing strategy is used again to transmit the messages to the destination node in the destination community. These schemes predict future encounter probability by using historical records. However, nodes' group identities are ignored. Xia et al. \cite{xia2015beeinfo} recently proposed an interest-based routing algorithm, called BEEINFO. This scheme is constructed based on a fact that people usually gather together to obtain and share their interest information. Therefore, they utilize interests to form communities, and design different forwarding strategies. However, they fail to protect the sensitive interest information.

The motivation for the proposed PRIF approach is to protect nodes' interests and make forwarding decisions. Specifically, to detect and maintain communities, similar to BEEINFO, PRIF uses interests to construct communities. To protect nodes' interests, a privacy-preserving authentication protocol is designed. Considering people's regular behaviors, PRIF gathers nodes' community information, and predicts the future destination community or destination node encounter probability.

\section{Privacy-Preserving Interest-Based Forwarding}
This section elaborates the design details of PRIF. We first give an overview of the whole scheme, and introduce the community detection method followed by the key concept of PRIF: a new social metric as \textit{community energy}. In addition, we powered our proposed work with efficient key management scheme based on some previous works \cite{du2009transactions,du2007effective,xiao2007survey}.Then, we introduce the privacy-preserving interest-based forwarding scheme.

\subsection{Overview}
The system model considered in this paper is a typical SIOV application scenario. There are potentially three kinds of mobile objects in the street: cars, buses, and pedestrians. Each car or bus owns a vehicle device, and each pedestrian carries a mobile device. They communicate with each other through wireless interface such as Wi-Fi or Bluetooth. In this paper, we use nodes to represent these devices. The aim of the application is to design an effective forwarding strategy by using these mobile nodes, without disclosing their interest information.

Since mobile nodes are used and controlled by people, the carriers\footnote{Carrier: Refers to the vehicle/individual carrying a mobile device. It does not represent the cellular service provider.} behavior can be an exact indicator of the nodes behavior. Thus, we take advantage of the human social property (i.e., interests) to make forwarding decisions. Normally, people have different interests. Although interests usually change over time, they can be considered relatively stable in a given time period. For example, some people are interested in reading in day-to-day activities, but during the World Cup, they may pay more attention to football. People with the same interest get together more often than others to obtain and share their interest information. For instance, people who share the interest of shopping appear frequently in the shopping malls but they nearly have no interaction with those who are interested in fishing. We assume a community is only related to one interest. If a person has more than one interest, he/she will belong to several communities simultaneously. In order to make our scheme easily understandable, similar to \cite{xia2015beeinfo}, each node is assumed to only hold one interest. During the forwarding process, node's interest information should be protected. We summarize the assumptions of the system below.
\begin{itemize}
	\item There are three types of mobile nodes (cars, buses and pedestrians) in the application scenario, which forms a typical SIOV.
	
	\item  There are no malicious nodes, and nodes are fully cooperative when forwarding messages.
	
	\item Each node only has one interest, and nodes with the same interest form a community.
	
	\item Each node must register itself with Trust Authority (TA).
	
\end{itemize}

The notations used in this paper are listed in Table \ref{t1}.

\begin{table}[tb]
	\caption{Notations}
	\begin{tabular}{ l|l}
		\hline
		\hline
		Symbol & Definition  \\
		\hline
		\hline
		M & Messages to deliver \\
		\hline
		$N_s$ & Source node \\
		$N_d$ & Destination node \\
		$N_i$ & Intermediate node \\
		\hline
		$I_s$ & Interest of the source node \\
		$I_d$ & Interest of the destination node \\
		$I_i$ & Interest of an intermediate node \\
		\hline
		$\mathbb{E}\_ \mathbb{I}_{(a,b)}$ & Inter-community energy between $a$ and $b$ \\
		$\mathbb{E}\_ \mathbb{C}_{(a,i)}$ & Intra-community energy between $a$ and the community $i$ \\
		\hline
		$\alpha$ & Inter-community energy prediction factor \\
		$\beta$ & Intra-community energy prediction factor \\
		\hline
		\hline
	\end{tabular}
	\label{t1}
\end{table}

\subsection{Community Energy}
In this section, we will introduce the concept of community energy which is inspired by molecular chemistry.
\subsubsection{Inter-community Energy}
In reality, molecules are composed of atoms and there exist forces among atoms. Similarly, we assume a force is generated when two nodes encounter one another. The force, called inter-community energy, represents their social tie and determines their contact strength. The stronger energy a node has, the more opportunities it has to successfully deliver messages. Note that the inter-community energy is only generated among nodes of the same community. We use Eq. \eqref{e1}, shown below, to define the inter-community energy between the nodes $a$ and $b$,
\begin{equation}
\begin{aligned}
\mathbb{E}\_\mathbb{I}_{(a,b)}(N) = \frac{d_{(a,b)}(N)}{t_{(N-1,N)}},
\end{aligned}
\label{e1}
\end{equation}

\noindent where $d_{(a,b)}(N)$ is the contact duration between $a$ and $b$ in the $N$-th encounter, and $t_{(N-1,N)}$ represents the duration that has elapsed from $(N-1)$-th encounter end to $N$-th encounter end.

\begin{figure*}[htb]
	\centering
	\subfigure[Inter-community energy]{
		\label{CE1} 
		\begin{minipage}[b]{0.28\textwidth}
			\centering
			\includegraphics[width=\textwidth]{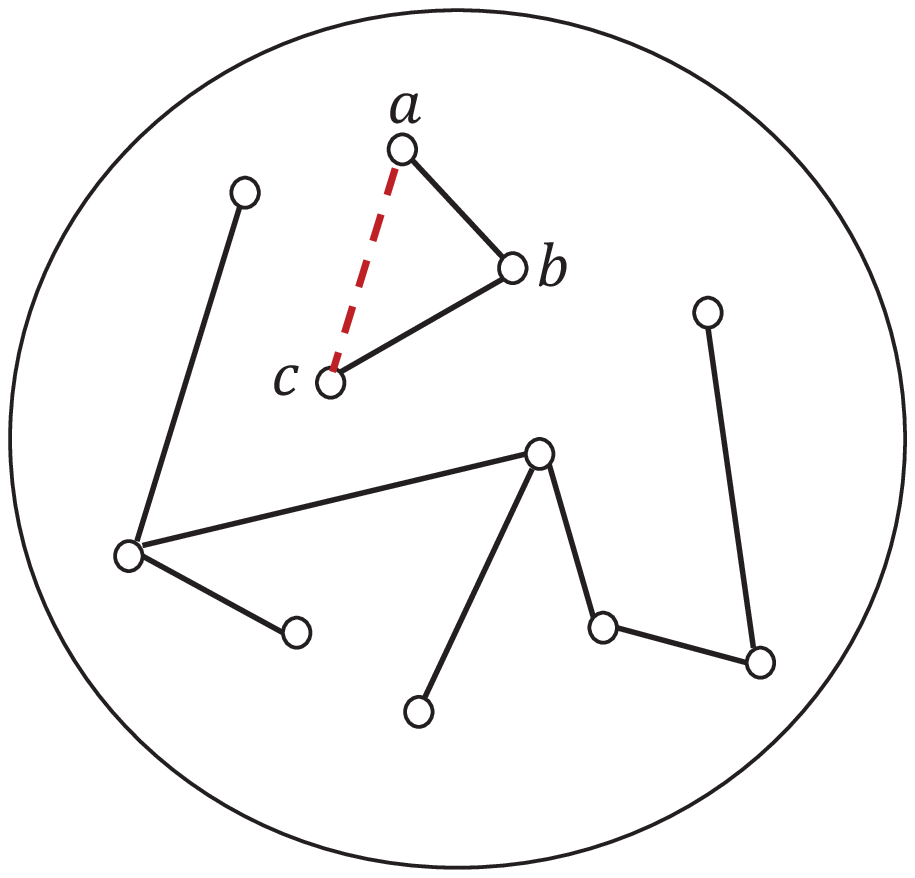}
	\end{minipage}}%
	~
	\centering
	\subfigure[Intra-community energy]{
		\label{CE2} 
		\begin{minipage}[b]{0.40\textwidth}
			\centering
			\includegraphics[width=\textwidth]{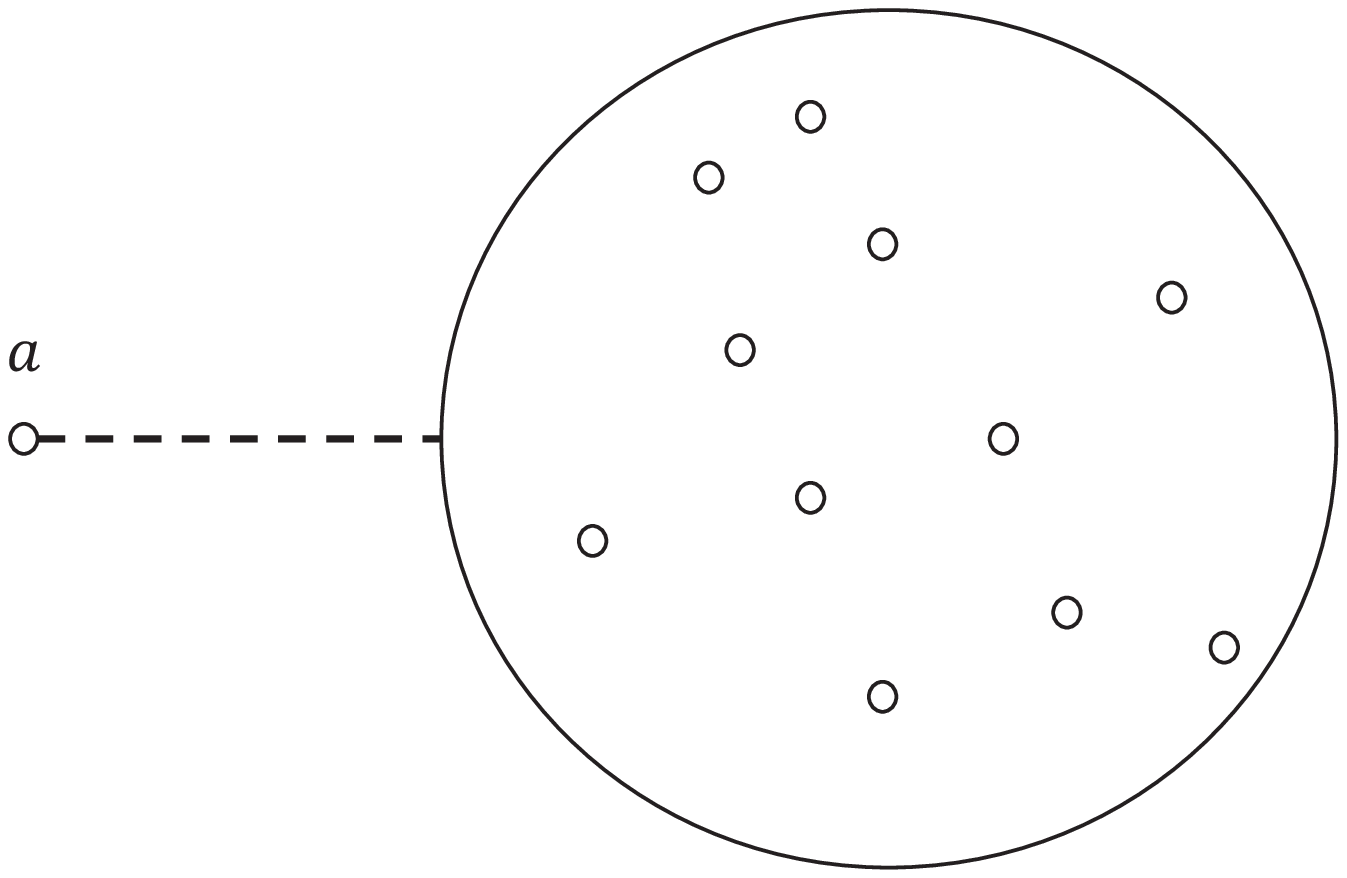}
	\end{minipage}}%
	\caption{Community energy.}\label{fig:bmbs}
	\vspace{-0.3cm}
\end{figure*}

The inter-community energy has a transitive property, which is based on an observation in reality. For example, if a person A frequently meets B, and meanwhile B frequently meets C, then A is also considered as a good forwarder to deliver messages to C though they may not encounter one another. Similarly, as shown in Fig. \ref{CE1}, a node $a$ establishes an energy to a node $b$, and $b$ builds an energy to a node $c$. Then, an indirect energy between $a$ and $c$ is generated as in Eq. \eqref{e2}, which is similar to \cite{PRoPHET},

\begin{equation}
\begin{aligned}
\mathbb{E}\_\mathbb{I}_{(a,c)}  = \mathbb{E}\_\mathbb{I}_{(a,c)(old)}  + (1 - \mathbb{E}\_\mathbb{I}_{(a,c)(old)}) \\
\times \mathbb{E}\_\mathbb{I}_{(a,b)} \times \mathbb{E}\_\mathbb{I}_{(b,c)}.
\end{aligned}
\label{e2}
\end{equation}

The nodes with high energy in the past usually are good forwarders in the future. Therefore, we define the energy prediction as in Eq. \eqref{e3} using an Exponential Weighted Moving Average (WEMA) \cite{grant},

\begin{equation}
\begin{aligned}
\mathbb{E}\_\mathbb{I}_{(a,b)}(N+1 )  = \alpha \times \mathbb{E}\_\mathbb{I}_{(a,b)}(N-1)
\\ 
+ (1- \alpha) \times \mathbb{E}\_\mathbb{I}_{(a,b)}(N),
\end{aligned}
\label{e3}
\end{equation}
where $\alpha$ is the inter-community energy prediction factor.

\subsubsection{Intra-community Energy}
In social networks, if a person encounters others from the same community frequently, the person can be considered as a good choice to forward messages destined for this community. We utilize degree centrality, which is the number of community nodes that a node encounters, to measure the community strength of the node, as shown in Fig. \ref{CE2}. However, considering the fast movement of mobile nodes, it may not be reasonable to directly use degree centrality. For example, a car may encounter many nodes interested in shopping around a shopping mall but fewer such nodes will be encounter after passing by the mall. Thus, we use average degree centrality to represent the intra-community energy between the node $a$ and the community $i$, as in Eq. \eqref{e4},

\begin{equation}
\begin{aligned}
\mathbb{E}\_ \mathbb{C}_{(a,i)}(N) = \frac{\sum^{k=N}_{k=1}n}{t_N},
\end{aligned}
\label{e4}
\end{equation}

In Eq. \eqref{e4}, $\sum^{k=N}_{k=1}n$ is the total number of nodes belonging to the same community $i$ that a node encounters from the first encounter to the $N$-th encounter, and $t_N$ is the duration time. If $a$ does not encounter members from the community $i$ for a long time, its intra-community energy $\mathbb{E}\_ \mathbb{C}_{(a,i)}$ will decrease sharply. In addition, we use Eq. \eqref{e5} to combine the past and present observations to predict the future intra-community energy. $\beta$ is the intra-community energy prediction factor, which is similar to $\alpha$ in Eq. \eqref{e3}.

\begin{equation}
\begin{aligned}
\mathbb{E}\_ \mathbb{C}_{(a,i)}(N+1)  = \beta \times \mathbb{E}\_ \mathbb{C}_{(a,i)}(N-1)
\\
+ (1-\beta) \times \mathbb{E}\_ \mathbb{C}_{(a,i)}(N).
\end{aligned}
\label{e5}
\end{equation}

\subsubsection{Energy Decay}

Finally, we consider the fact that if nodes do not encounter each other in a period of time, they may not remain good forwarders for each other. Thus, an evaporation/aging process is necessary. We use Eq. \eqref{e6} and Eq. \eqref{e7} to decay the community energy,

\begin{equation}
\begin{aligned}
\mathbb{E}\_ \mathbb{I}_{new} = \mathbb{E}\_ \mathbb{I}_{old} \times \gamma^k,
\end{aligned}
\label{e6}
\end{equation}

\begin{equation}
\begin{aligned}
\mathbb{E}\_ \mathbb{C}_{new} = \mathbb{E}\_ \mathbb{C}_{old} \times \gamma^k,
\end{aligned}
\label{e7}
\end{equation}
where $\gamma$ is the aging factor, and $k$ is the number of time intervals since the last time energy was aged.

When nodes move around, they share and gather interest information, and further update the above community energy information.

\subsection{Privacy-Preserving Interest-Based Forwarding}
In this section, we introduce the privacy-preserving interest-based forwarding scheme including system initialization, privacy-preserving authentication, forwarding process, message scheduling and buffer management strategies.
\subsubsection{System Initialization}
Let $p$ be a large prime, $\alpha \in Z^{*}_{p}$, and the order of $\alpha$  be  $q$, where $q$ is a large prime factor of $p-1$.   $H_{1} : \{0, 1\}^{*}\rightarrow
Z^{*}_{q}$ and  $H_2 : \{0, 1\}^{*}\rightarrow
\{0, 1\}^{\kappa}$ are cryptographic hash
functions.

TA generates    a  certificate revocation list $\mathcal{RL}$,  which is originally empty and public. In order to create the group $\mathcal{G}_l$ (i.e., the community $\mathcal{G}_l$), $l\in [1,L],$ TA randomly chooses $a_l \in  Z^{*}_{q}$ and computes $y_l=\alpha^{a_l}$ mod $p$. Then, TA sets the group secret key $msk$ for  $\mathcal{G}_l$ as $a_l$. In addition, TA generates a group ID $GID_l$ for  $\mathcal{G}_l$.

When a node $U_i$ wants to join the group $\mathcal{G}_l$,  TA registers  it. TA generates a certificate   and sends it to $U_i$ over an authenticated private channel. TA   randomly selects    a string $id_i$ and  $k_{i} \in Z^{*}_{q}$, and generates a Schnorr signature
$\sigma_i=(e_i, s_i)$, where $e_i = H_1(id_i, \alpha^{k_i}$ mod $p$)
and $s_i = {a_l}e_i + k_i$ mod $q$.  $U_i$'s certificate is
$\mathrm{\tt{cert}}_i = (id_i, e_i, s_i, y_l)$.  Note that, $y_l$ is known by  the members of all the groups and TA, while the certificate $\mathrm{\tt{cert}}_i$ is only known by $U_i$ itself. If $U_i$ wants to leave the group, TA inserts $id_i$ into $\mathcal{RL}$.

\subsubsection{Privacy-preserving Authentication}
Assume that $U_i$ claims that it is affiliated to the group $\mathcal{G}_l$, and $U_j$ claims that it belongs in the group $\mathcal{G}_z$. After executing privacy-preserving authentication,    $U_i$ can identify if   $U_j$ belongs in $\mathcal{G}_z$ and   $U_j$ can identify if   $U_i$ is affiliated to $\mathcal{G}_l$. If they are in the same group,  we can conclude that they have the same interest.

We assume that a node $U_i$ with $(id_i,e_i,s_i,y_l)$ belongs to $\mathcal{G}_l$, and $\mathcal{G}_l$'s group ID is GID$_l$.  Assume $U_i$ encounters another node $U_j$, where $U_j$ claims that it is affiliated to $\mathcal{G}_z$. $U_i$ can communicate with $U_j$ to check if $U_j$ is affiliated to $\mathcal{G}_z$. Specifically, $U_i$ performs the following steps:

\begin{itemize}
	\item $U_i$ randomly selects $b_{i} \in \mathbb{Z}_{q}^{*}$. Here, $b_{i}$ mod $ q \neq 0$.
	
	\item $U_{i}$ calculates   ($Y_{i} = \alpha^{s_{i}} \cdot {y_l}^{-e_{i}}$ mod $p$) =
	($\alpha^{k_{i}}$ mod $p$), and $B_{i} = \alpha^{b_{i}}$ mod $p$.
	
	\item $U_{i}$ sends $M_i=(GID_l, id_i, Y_{i}, B_{i})$ to $U_j$.
\end{itemize}

Similarly, $U_j$ generates $M_j=(GID_z, id_j, Y_{j}, B_{j})$, and sends it to $U_i$.

If   $id_{j}$   is not listed in $\mathcal{RL}$ and $(Y_{j})^{(p-1)/q}$ $\not\in$ $\{0, 1\}$,
$ {U}_{i}$   computes $K_{i,j}
=  B_{j} ^{s_{i}}$ mod $p$ and
sets $v_{i}=(h_{i,j},
sid_{i}$), where $h_{i,j} =
H_2(K_{i,j},  sid_{i})$ and $sid_{i} =
[M_{i}||M_{j}]$.
Otherwise,  $ {U}_{i}$
randomly selects $h_{j}'\leftarrow \{0, 1\}^{\kappa}$, then sets
$v_{i}=(h_{j}'$, $sid_{i})$ and  $reject = T$. $U_{i}$ sends $v_{i}$ to $U_j$.  Similarly, $U_{j}$ sends $v_{j}=(h_{j,i},sid_j)$ to $U_i$, where
\begin{equation} \notag
\begin{aligned}
h_{j,i} &= H_2(K_{j,i} \ \mathrm{mod} \ p, sid_j)  \\
         & = H_2(B_i^{s_j} \ \mathrm{mod} \ p, sid_j) \\
         &= H_2(\alpha^{b_i \cdot (a_ze_j+k_j)} \ \mathrm{mod} \ p, sid_j)
\end{aligned}
\end{equation}

If $reject = T$, then $ {U}_{i}$ rejects communication. Otherwise, $ {U}_{i}$ performs the following steps:

\begin{itemize}
	\item
	
	After ${U}_{i}$ receives $v_{j}$,
	${U}_{i}$ computes $h_{j,i}' =  H_2({({y_z}^{H_1(id_j,Y_j)}Y_j)}^{b_i}$ mod $p, sid_{j})$.
	\item
	$U_i$ checks whether $h_{j,i}'$ equals to $h_{j,i}$. Since 
	\begin{equation} \notag
	\begin{aligned}
	h_{j,i}' &= H_2({({y_z}^{H_1(id_j,Y_j)}Y_j)}^{b_i} \ \mathrm{mod} \ p, sid_{j})  \\
	          &= H_2(({y_z}^{e_j}Y_j)^{b_i} \  \mathrm{mod} \ p, sid_j) \\
	          &= H_2(\alpha^{a_ze_j} \alpha^{k_j})^{b_i} \ \mathrm{mod} \ p, sid_j) \\
	          &= H_2(\alpha^{(a_ze_j+k_j)b_i } \ \mathrm{mod} \ p, sid_j)
	\end{aligned}
	\end{equation}
	
	Thus, if   $h_{j,i}'\neq h_{j,i}$,
	$ {U}_{i}$ can conclude  that $U_{j}$ is an invalid
	participant. Otherwise, $ {U}_{i}$ can conclude that the group ID  of $ {U}_{j}$ is GID$_z$.
	
\end{itemize}

According to GID$_z$, $ {U}_{i}$ can conclude whether $ {U}_{j}$  belongs in the same group with $ {U}_{i}$.

\subsubsection{Forwarding Process}
When mobile nodes are in communication range, they will communicate with each other. The forwarding process consists of two parts: community energy awareness and message forwarding strategy.

\noindent\textbf{Community energy awareness.}
When two nodes (for example $N_s$ and $N_i$) encounter, they first check if they are in the same community in a privacy-preserving way, and then update the community energy. If they are affiliated to different communities, they accumulate the community number and update their intra-community energy. Otherwise, they compute the connection time and update their inter-community energy. We give the pseudocode of community energy awareness in Algorithm 1.

\begin{algorithm}[!h]
	\label{A1}
	\SetCommentSty{small}
	\LinesNumbered
	\caption{Pseudocode for community energy awareness}
	\For{ all intermediate nodes $N_i$ connected to $N_s$}{
		\If {$I_i == I_s$} {
			$//$ Update the direct inter-community energy; \\
			\If {$N_s$ has the inter-community energy record of $N_i$}{
				Update the inter-community energy with Eq. \eqref{e1}; \\}
			\Else{
				Initialize the inter-community energy between $N_i$ and $N_s$;
			}
			$//$ Update the indirect inter-community energy;  \\
			\For {all connected nodes of $N_i$ and $N_s$}{
				Update the indirect inter-community energy with Eq. \eqref{e2};
			}
		}
		\Else{
			$//$ Update intra-community energy; \\
			\If {$N_s$ has intra-community energy record of $N_i$}{
				Update intra-community energy with Eq. \eqref{e4};
			}
			\Else{
				Initialize intra-community energy;
			}
		}
		Predict inter-community energy with Eq. \eqref{e3}; \\
		Predict intra-community energy with Eq. \eqref{e5}; \\
	}
\end{algorithm}

\noindent\textbf{Message forwarding strategy.}
The message forwarding strategy is the core of PRIF. By using community energy, the best forwarders can be chosen for the destination. According to the communities of $N_s$, $N_i$ and $N_d$, PRIF uses different message forwarding strategies.

Assume that a node $N_s$ with a message $M$ destined for $N_d$ meets another node $N_i$. If $N_i$ is not the destination node and $N_s$, $N_i$ and $N_d$ belong in the same community, inter-community energy will be used to make forwarding decisions. If $N_i$ has higher inter-community energy to the destination, it will be selected as a better forwarder. Otherwise, $N_s$ will stop forwarding and wait for a better opportunity. If $N_s$ does not share the same interest with $N_d$, then there are only two cases where the forwarding process can occur: (1) $I_i == I_d$. In this case, $N_i$ belongs to the destination community; (2) $N_i$ does not belong to $N_d$'s community and $\mathbb{E}\_ \mathbb{C}_{(N_s,I_d)} < \mathbb{E}\_ \mathbb{C}_{(N_i,I_d)}$. Otherwise, $N_s$ will continue holding the message $M$.

As a whole, PRIF looks for active intermediate nodes (with higher inter-community or intra-community energy) which will allow fast transfer of message to the destination node or destination community. When the message has reached the destination, it broadcasts a response message to inform all nodes which still maintain the message to discard it. We give the pseudocode of message forwarding strategy in Algorithm \ref{A2}. In Algorithm 2, $E$ denotes a secure identity-based encryption algorithm \cite{BonehF01}, $ID_{d}$ presents the pseudo identity of $N_d$, and $E_{ID_{d}}(M)$ is the ciphertext of the message $M$.

\begin{algorithm}[htb]
	\label{A2}
	\SetCommentSty{small}
	\LinesNumbered
	\caption{Pseudocode for message forwarding strategy}
	When $N_s$ with a  message $M$ destined for $N_d$ encounters a node $N_i$. \\
	\If{$N_i$ is $N_d$}{
		Deliver $E_{ID_{d}}(M)$ from $N_s$ to $N_d$;
	}
	\Else{
		\If {$I_s$ == $I_d$}{
			$//$ $N_s$ belongs to the destination community; \\
			\If {$I_i$ == $I_d$}{
				$//$ $N_i$ belongs to the destination community; \\
				\If{$\mathbb {E} \_ \mathbb{I}_{(N_s,N_d)} < \mathbb{E}\_ \mathbb{I}_{(N_i,N_d)}$}{
					$//$ $N_i$ has higher inter-community energy; \\
					Deliver $E_{ID_{d}}(M)$ from $N_s$ to $N_i$; \\
				}
			}
		}
		\Else {
			$//$ $N_s$ does not belong to the destination community; \\
			\If{$I_i$ == $I_d$}{
				$//$ $N_i$ belongs to the destination community; \\
				Deliver $E_{ID_{d}}(M)$ from $N_s$ to $N_d$;  \\
			}
			\Else{
				$//$ $N_i$ does not belong to the destination community; \\
				\If{$\mathbb{E}\_ \mathbb{C}_{(N_s,I_d)} < \mathbb{E}\_ \mathbb{C}_{(N_i,I_d)}$}{
					$//$ $N_i$ has higher intra-community energy; \\
					Deliver $E_{ID_{d}}(M)$ from $N_s$ to $N_i$; \\
				}
			}
		}
	}
\end{algorithm}

\subsubsection{Message Scheduling and Buffer Management}
Since all mobile nodes have limited resources (i.e., battery power and buffer size), it is necessary to design message scheduling and buffer management strategies to improve the forwarding efficiency. The message scheduling policy decides in what order to deliver messages so as to ensure messages can be delivered to the destination node with higher delivery opportunities. In PRIF, we design strategies based on their communities. The message whose interest is the same as that of the current node will have a priority. When the buffer size reaches its capacity, the buffer management strategy decides which messages will be discarded if new messages arrive. Moreover, similar to message scheduling, the buffer management scheme is also based on communities. Details of both are given below.

\noindent\textbf{Message scheduling algorithm.}
When  $N_i$ is selected as a message forwarder and it has a set of messages to be delivered, then the relation between $N_i$ and $N_d$ is a major factor that needs to be considered. Specifically, the algorithm orders messages with the following priority rules: (1) the messages satisfying $I_i == I_d$ will have priority. The messages satisfying this condition will be ordered according to inter-community energy. If the inter-community energy is equal, the newer message will be transmitted first; (2) for the messages which do not satisfy $I_i == I_d$, it suits intra-community transmission, hence intra-community energy of $N_i$ is considered. The messages with higher intra-community energy will have higher priority. If intra-community energy is equal, then the newer one will be transmitted first.

\noindent\textbf{Buffer management algorithm.}
The buffer management algorithm relies on the relation between the source node $N_s$ and the message. It discards the messages following the reverse order as  that of the message scheduling sequence: (1) the messages which have different interests with the destination nodes will be discarded first. In this condition, the messages with lower intra-community energy will be replaced first. In the case that intra-community energy is equal, the older one will be discarded; (2) we then consider the messages in which $I_s == I_d$. The messages with lower inter-community energy will be replaced first. If the intra-community energy is equal, the message coming later will be discarded.

\section{Security Analysis}

In this section, we will introduce the security model and prove that our scheme is privacy-preserving by showing that the attacks studied in \cite{wu2014security,wu2014mobifish,huang2014achieving} cannot be used to determine the participants' interest information.

\subsection{Security Model}

In this security model,  the privacy property is    defined by using a game between an adversary $\mathcal{A}$ and
a challenger $\mathcal{C}$.
The adversary $\mathcal{A}$'s goal   is to
learn about the players'
interest information. The adversary cannot learn the interest information unless
it can   distinguish the two executions: One where the challenger
$\mathcal{C}$ executes the protocols   as  honest
players, while the other where the adversary $\mathcal{A}$  runs the protocol with a
simulator.

Firstly, the challenger $\mathcal{C}$ creates  a group in which  $m$
members are included. Specifically, $\mathcal{C}$ generates $msk$   of the group. Besides, $\mathcal{C}$  generates  a certificate
$\mathrm{\tt cert}_{i}$ for the user $U_i$, where $i \in [1,m]$. Then,  $\mathcal{C}$
chooses  corrupted players and gives the certificates of the corrupted players to $\mathcal{A}$.  Subsequently,
$\mathcal{C}$ updates  $\mathcal{RL}$.

Afterwards,  $\mathcal{A}$    sends a polynomial number of
$\mathrm{\tt Start}(\Pi _{i}^{s}, G$), $\mathrm{{\tt Send}}(\Pi
_{i}^{s}, \Delta$),   and $\mathrm{\tt Corrupt}(
{U}_{i}$) queries  adaptively.  $\mathcal{C}$   picks at random a bit $b$ uniformly, where $b
\in \{0, 1\}$.  If $b = 1$, $\mathcal{C}$ answers
$\mathcal{A}$'s requests   as   honest players. Otherwise, $\mathcal{C}$
responds to $\mathcal{A}$ by using  the simulator. When $b=0$, $\mathcal{C}$
replies to the queries as follows:

\begin{itemize}

	\item
	
	$\mathrm{\tt Start}(\Pi _{i}^{s} $) and
	$\mathrm{{\tt Send}}(\Pi _{i}^{s},  \Delta$) queries:   $\mathcal{C}$  answers the queries with the messages
	generated by the simulator. $\mathcal{C}$ will set  $reject$ as $T$ and return $null$,  if $\Delta$ is  incorrect.

	\medskip

	\item
	$\mathrm{\tt Corrupt}({U}_{i}$):  $\mathcal{C}$ gives $ \mathrm{\tt cert}
	_{i}$ to $\mathcal{A}$ and updates the  revocation list.

\end{itemize}

Finally, $\mathcal{A}$ outputs a bit $b'$. $\mathcal{A}$ wins the game, if $b'= b$ holds. The advantage with which $\mathcal{A}$ wins the game is
defined to be
$$\mathrm{\tt Adv}  (\mathcal{A})=\big|2\cdot{\mathrm{Pr}}[b=b']-1 \big |.$$

\begin{definition}   The proposed protocol is said to be  privacy-preserving,
	if for  any probabilistic polynomial time
	adversary $\mathcal{A}$,    $\mathrm{\tt Adv}(\mathcal{A})$
	is   negligible.
\end{definition}

\subsection{Security Proof}

\begin{theorem}
	Assume $\mathcal{A}$ can ask the $H_{2}$ random oracle at most $q_{H_{2}}$ times. The proposed scheme is secure with the probability $1-2q_{H_{2}}/(q-1)$, where $q$ is the order of group $Z_{q}^{*}$ and $q$  is a large prime.
\end{theorem}

\noindent{\itshape{Proof}}. In order to prove that our protocol
is privacy-preserving, we design two games Game0 and Game1, where  Game1 denotes a simulation, and Game0 represents the real game.

In order to show that the adversary $\mathcal{A}$ can not distinguish its view in Game1 and its view  in
Game0,   Game1 is constructed as
follows.

\noindent\textbf{Simulation}. Assume
pid$_{i}^{s}$=$\{U_{i}, U_{j}\}$.

\begin{itemize}
	\item

	$\mathrm{\tt Start}(\Pi _{i}^{s}$):  $\mathcal{C}$ randomly chooses
	$id_i$ corresponding to $U_i$,  selects   randomly
	$k \in Z_{q}^{*}$  and calculates $Y_{i}  = g^{k}$ mod $p$, then picks
	randomly  $b_{i} \in \mathbb{Z}_{q}^{*}$ and calculates $B_{i}
	= \alpha^{b_{i}}$ mod $p$.  $\mathcal{C}$  replies with  $M_i=(id_i,
	Y_{i}, B_{i})$.
	
	\medskip
	\item
	$\mathrm{{\tt Send}}(\Pi _{i}^{s},  \Delta$):
	$\mathcal{C}$  randomly selects $h_{j}'\leftarrow \{0,
	1\}^{\kappa}$, then
	sets $v_{i}=(h_{j}', sid_{i}^{s})$.
	Output $v_{i}$.

	\item

	$\mathrm{\tt Corrupt}(\mathcal{U}_{i}$): $\mathcal{C}$ gives $ \mathrm{cert}
	_{i}  $ to $\mathcal{A}$ and inserts $id_{i}$
	to the revocation list.
\end{itemize}

For   any $\Pi^{s} _{i}$ and any $j \in \mathcal{D}$,
$ K_{i,j}$ and $ K_{j,i}$ is defined via the messages  $(id_i,
Y_{i}, B_{i})$ and   $(id_j, Y_{j}, B_{j})$, where $(id_i, Y_{i},
B_{i})$ is sent by $\Pi^{s} _{i}$ of
$U_{i}$   on that session. $(id_j, Y_{j}, B_{j})$ is sent by
$\mathcal{A}$ to $\Pi^{s}_{i}$. That is, $K_{i,j} = (y^{H(id_j,Y_j)}Y_j)^{b_i}$ mod $p,$ and $K_{j,i} = (y^{H(id_i,Y_i)}Y_i)^{b_j}$ mod $p$. Let
$\mathcal{E}_{H_{2}}(i)$ denote the event that $\mathcal{A}$
sends $H_{2}$ query on $( K_{i,j}, sid_{i}^{s} )$ or $( K_{j,i}, sid_{i}^{s})$.

We can observe the difference between the two games. That is, $h_{k}'$ is randomly selected in Game1, where
$k \in \mathcal{D}$. Thus,
$\mathcal{A}$
can not distinguish its view  in Game1 from its view in Game0
unless
$\mathcal{E}_{H_{2}}(i)$ happens. In
the proposed  scheme,
$K_{i,j} = (y^{H(id_j,Y_j)}Y_j)^{b_i}$ mod $p,$  $K_{j,i} = (y^{H(id_i,Y_i)}Y_i)^{b_j}$ mod $p$ and  $\mathcal{A}$   cannot know  $y$. Therefore,  $K_{i,j}$ and   $K_{j,i}$ can be considered as $\alpha ^{x_{1}}$ mod $p$  and $\alpha ^{x_{2}}$ mod $p$  respectively, for some  unknown $x_{1} \in \mathbb{Z}_{q}^{*}$ and some  unknown   $x_{2} \in \mathbb{Z}_{q}^{*}$ for $\mathcal{A}$. Then,    the probability with which $\mathcal{A}$ can query $K_{i,j}$ or   $K_{j,i}$ is  $2 q_{H_{2}}/(q-1)$  at most.
That is, the probability with which  event $\mathcal{E}_{H_{2}}(i)$ happens is $2 q_{H_{2}}/(q-1)$ at most. Moreover, $\mathcal{A}$
can   distinguish its view   in Game1 from that in Game0 with $2
q_{H_{2}}/(q-1)$ at most.   $2q_{H_{2}}/(q-1)$ becomes negligible, since $q$ is a large prime. That is, $\mathcal{A}$ can not
distinguish its view in Game1 from that in Game0. Therefore, our
protocol captures the privacy-preserving property.

\section{Performance Evaluation}
We have conducted extensive experiments to evaluate the performance of the proposed PRIF and compared it with the following routing and forwarding methods, i.e., BEEINFO-D$\&$S \cite{xia2015beeinfo}, Epidemic \cite{epidemic}, and PRoPHET \cite{PRoPHET}.

Following metrics have been used for performance comparison:
\begin{itemize}
	\item Delivery ratio: the average ratio of successfully delivered messages to all created messages from the sources to the destinations.
	\item Overhead: the percentage of relayed messages which excludes the delivered messages.
	\item Average hop count: the average number of hops when messages are delivered successfully.
\end{itemize}

\subsection{Simulation Settings}
Similar to BEEINFO-D$\&$S \cite{xia2015beeinfo} and other DTN routing schemes such as \cite{grant}, \cite{ONE-1}, \cite{ONE-2} and \cite{ONE-3}, The Opportunistic Network Environment (ONE) Simulator \cite{ONE} is used to evaluate the performance of the PRIF.

In our experiments, five groups of nodes are considered, including two pedestrian groups, two car groups, and one bus group. All groups consist of 40 nodes except the bus group which has 6 nodes. There are two kinds of Bluetooth interface to realize wireless transmission: one is used for cars and pedestrians where the communication range is 10 m and transmission rate is 2 Mb/s, and the other one is used for buses with a higher communication range and transmission rate (i.e., 100 m and 10 Mb/s). Messages are only generated by nodes of cars and pedestrians groups, every 50-90 s. The size of message is set as 0.5-1 MB. We implement the experiments by varying the values of two important factors: the buffer size (10-50 MB) and TTL (600-3600 min). Detailed simulation parameters are listed in Table \ref{t2}.

\begin{table} [tb]
	\caption{Parameters used in evaluation.}
	\begin{tabular}{ l|l}
		\hline
		\hline
		Parameter & Value or Range \\
		\hline
		\hline
		Simulation time & 400000 s \\
		Time window & 30 s \\
		Warm up time & 5000 s \\
		Area & 4500 $\times$ 3400 m$^2$ \\
		Speed of pedestrians & $0.5 \sim 1.5$ m/s \\
		Speed of cars & $2.7 \sim 13.9$ m/s \\
		Speed of buses & $7 \sim 10$ m/s \\
		Wait time at destination & $100 \sim 200$ s\\
		Message TTL & 600 min \\
		Event interval & $50 \sim 90$ s \\
		Message size & $500 \sim 1024$ KB \\
		Number of nodes in each car/pedestrian group & 40 \\
		Number of nodes in each bus group & 6 \\
		$\alpha, \beta$ & 0.3 \\
		$\gamma$ & 0.98 \\
		\hline
		\hline
	\end{tabular}
	\label{t2}
\end{table}

\subsection{Simulation results and analysis}
\begin{figure*}[htb]
	\centering
	\subfigure[]
	{ \includegraphics[width=.30\textwidth]{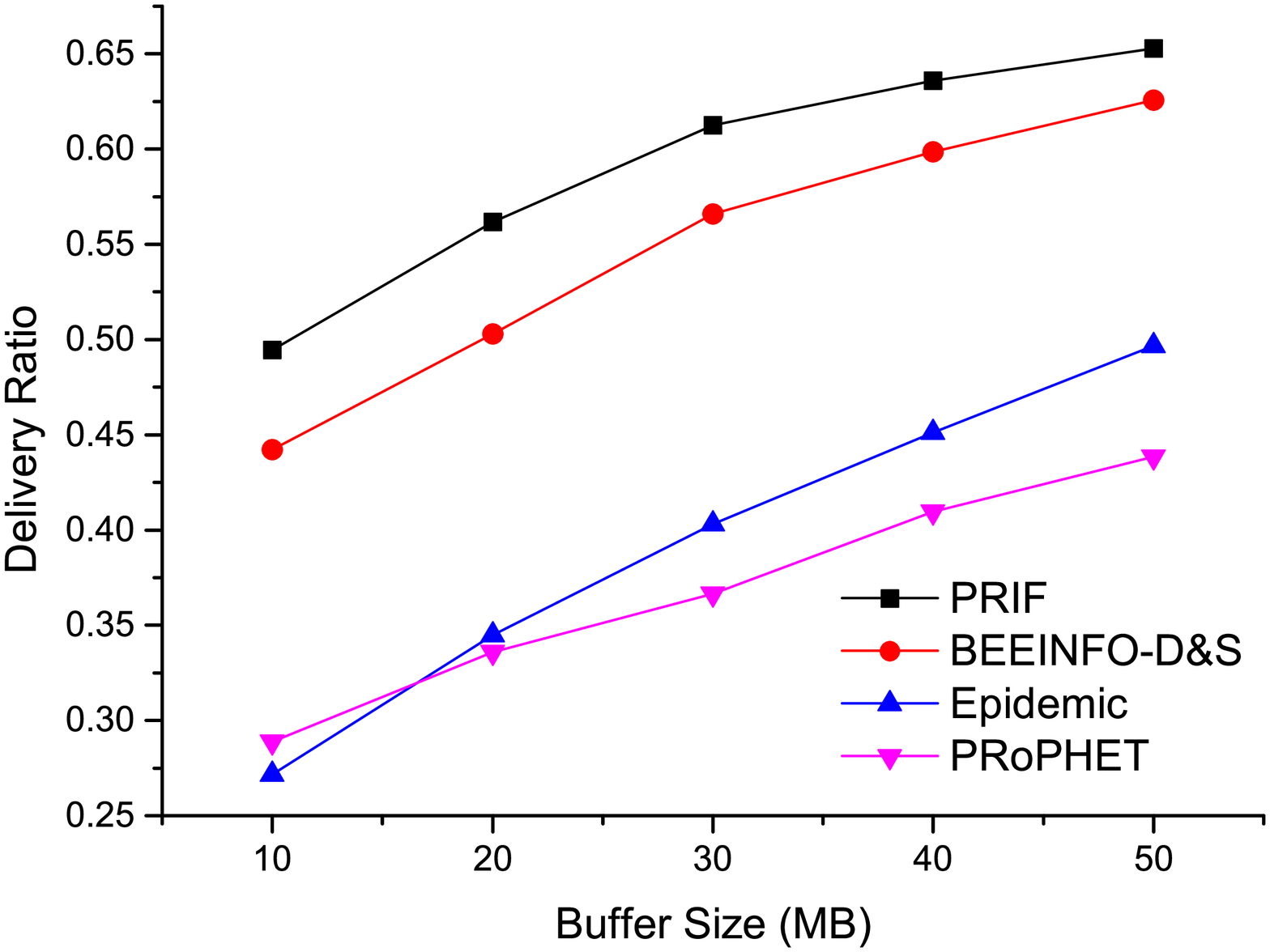}}
	~~
	\subfigure[]
	{\includegraphics[width=.30\textwidth]{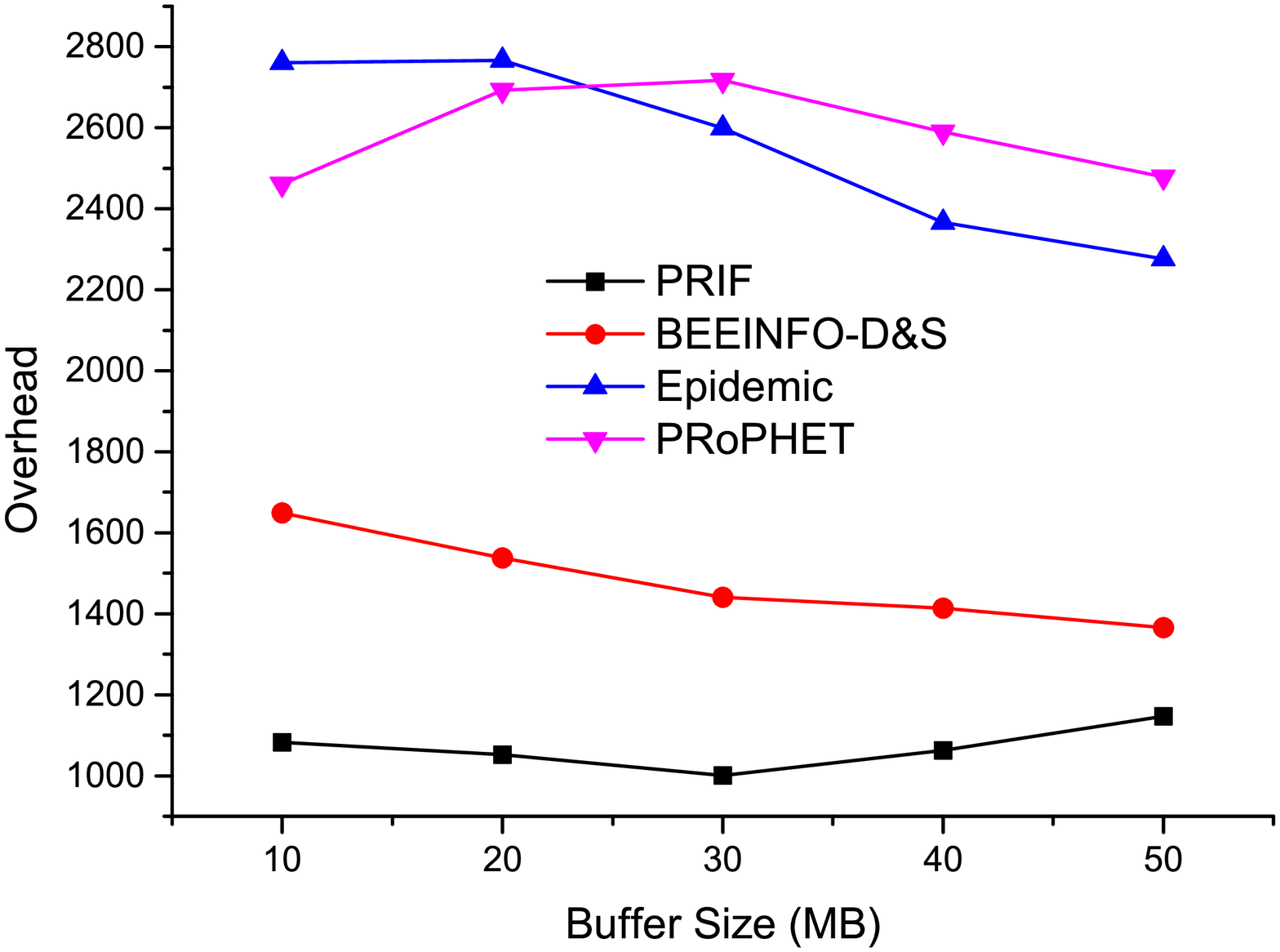}}
	~~
	\subfigure[]
	{\includegraphics[width=.30\textwidth]{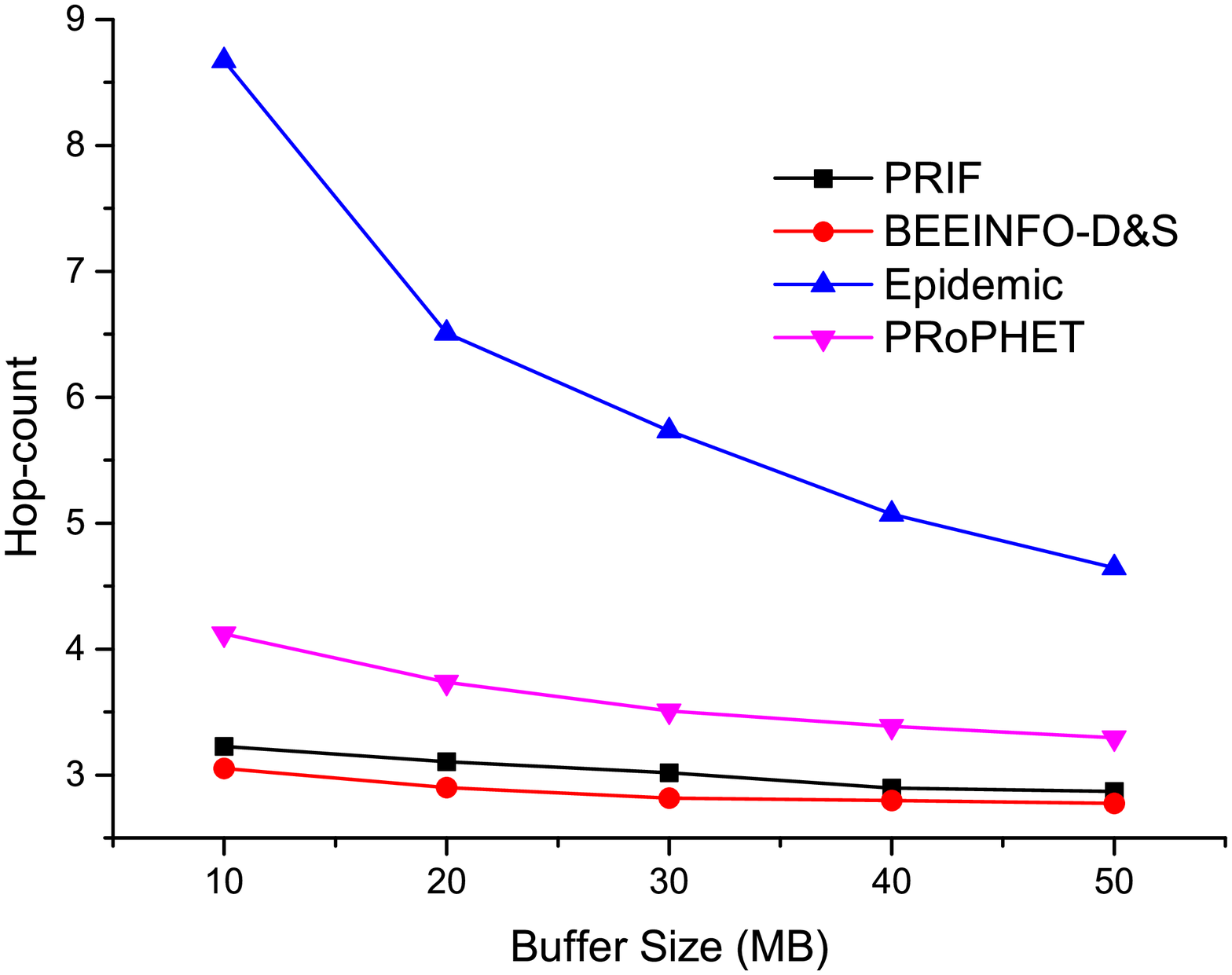}}
	~~
	\caption{Performance over buffer size.}
	\label{fig:BS}
\end{figure*}

\begin{figure*}[htb]
	\centering
	\subfigure[]
	{ \includegraphics[width=.30\textwidth]{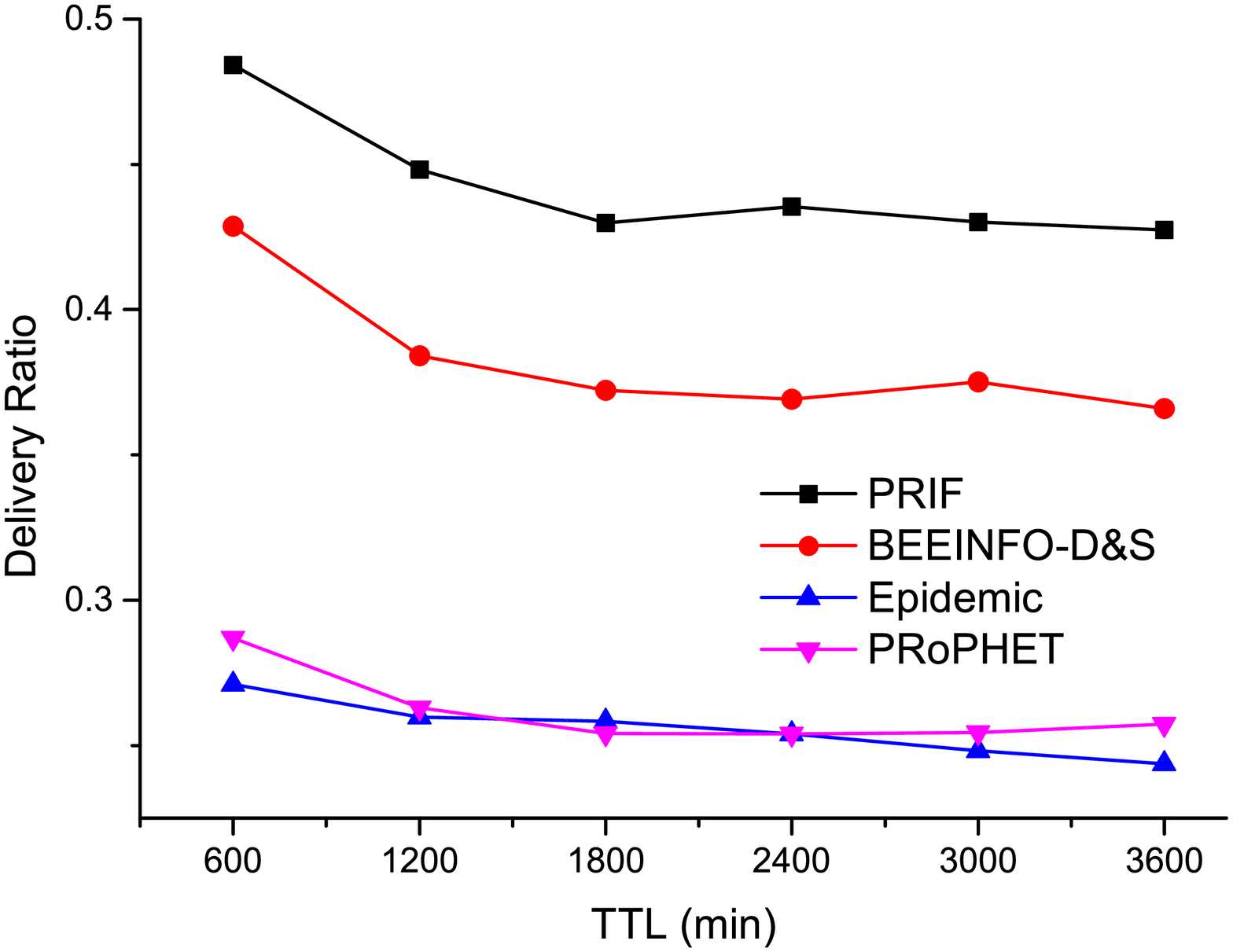}}
	~~
	\subfigure[]
	{\includegraphics[width=.30\textwidth]{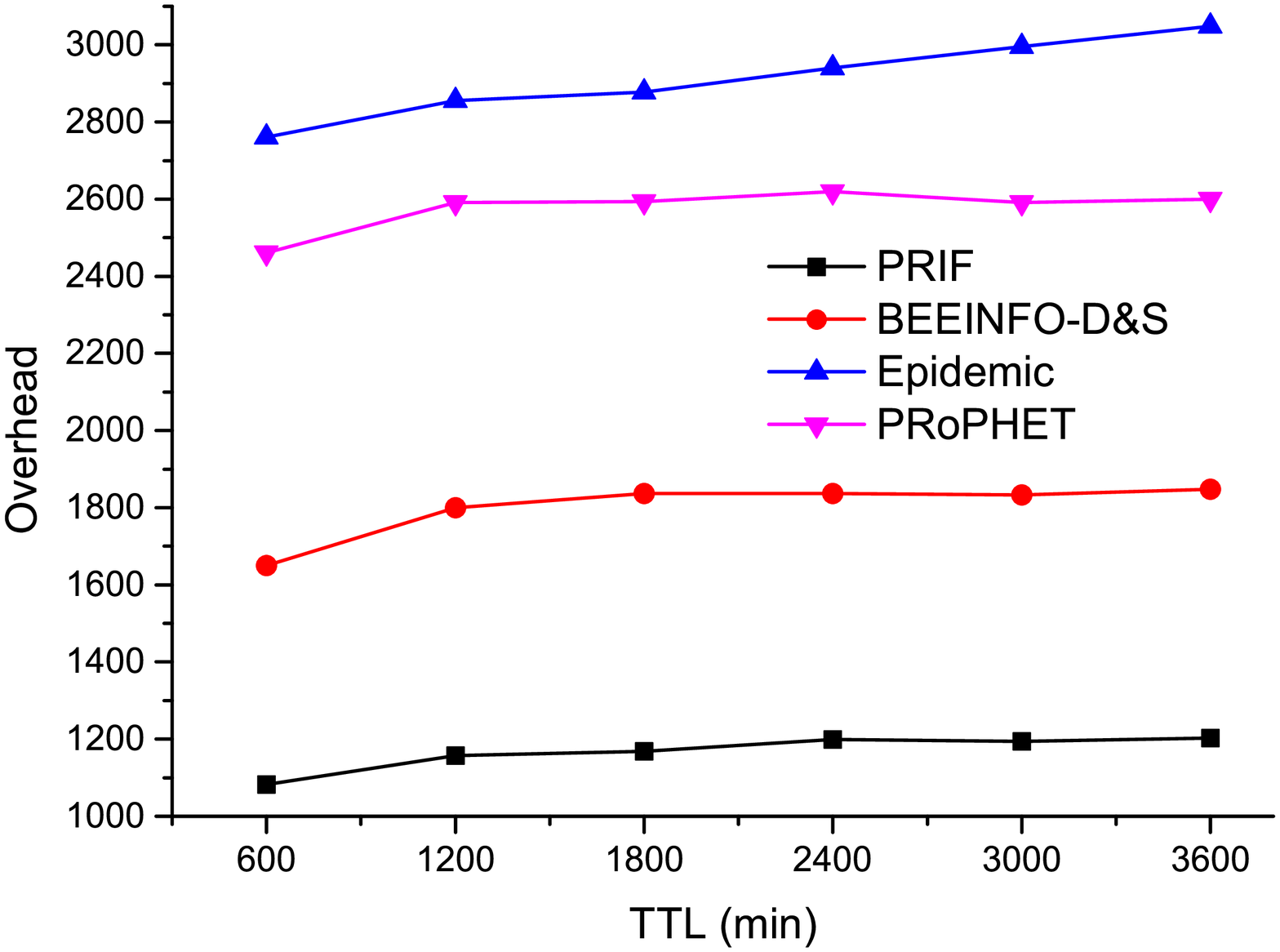}}
	~~
	\subfigure[]
	{\includegraphics[width=.30\textwidth]{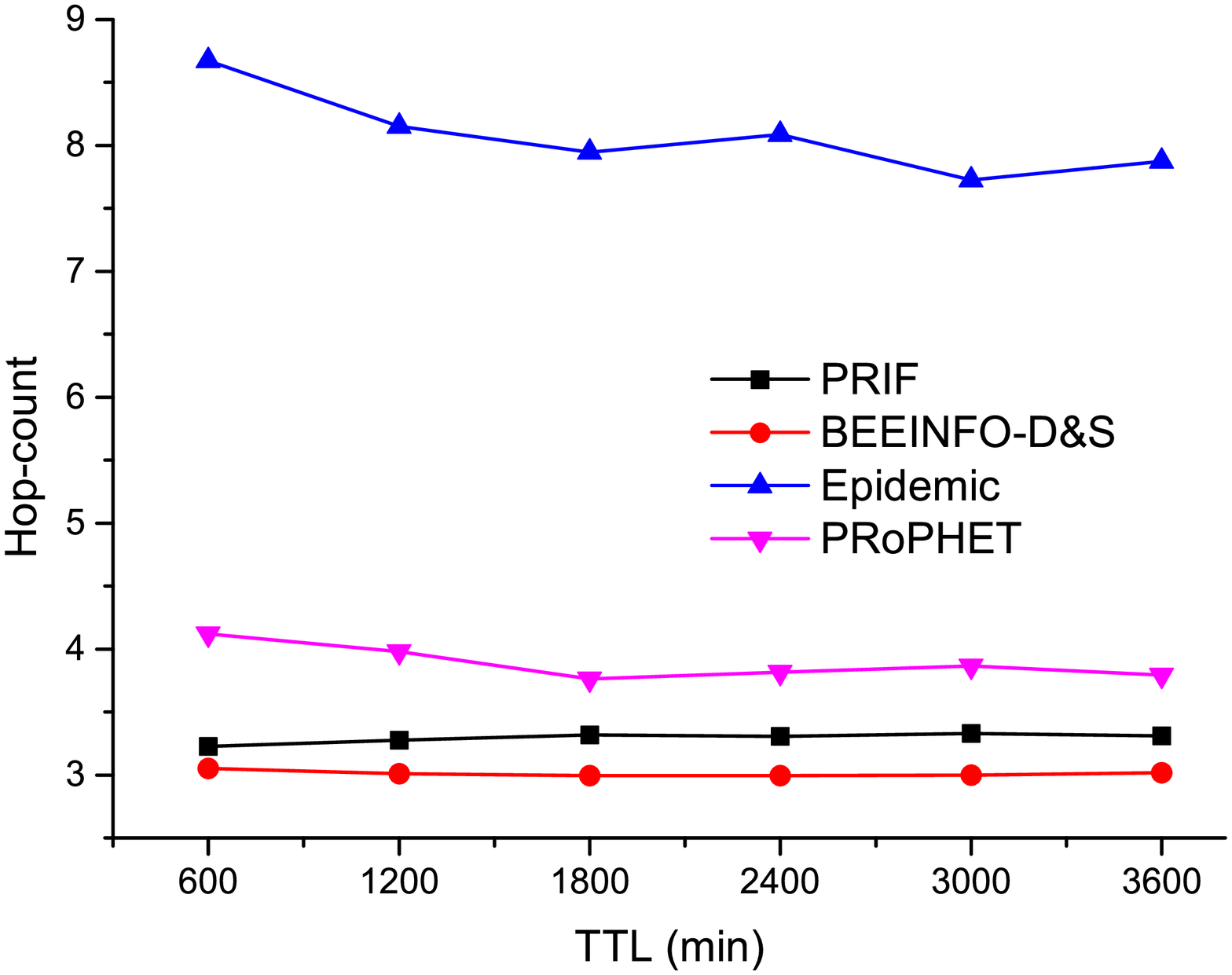}}
	~~
	\caption{Performance over TTL.}
	\label{fig:TTL}
\end{figure*}

\begin{figure*}[htb]
	\centering
	\subfigure[]
	{ \includegraphics[width=.30\textwidth]{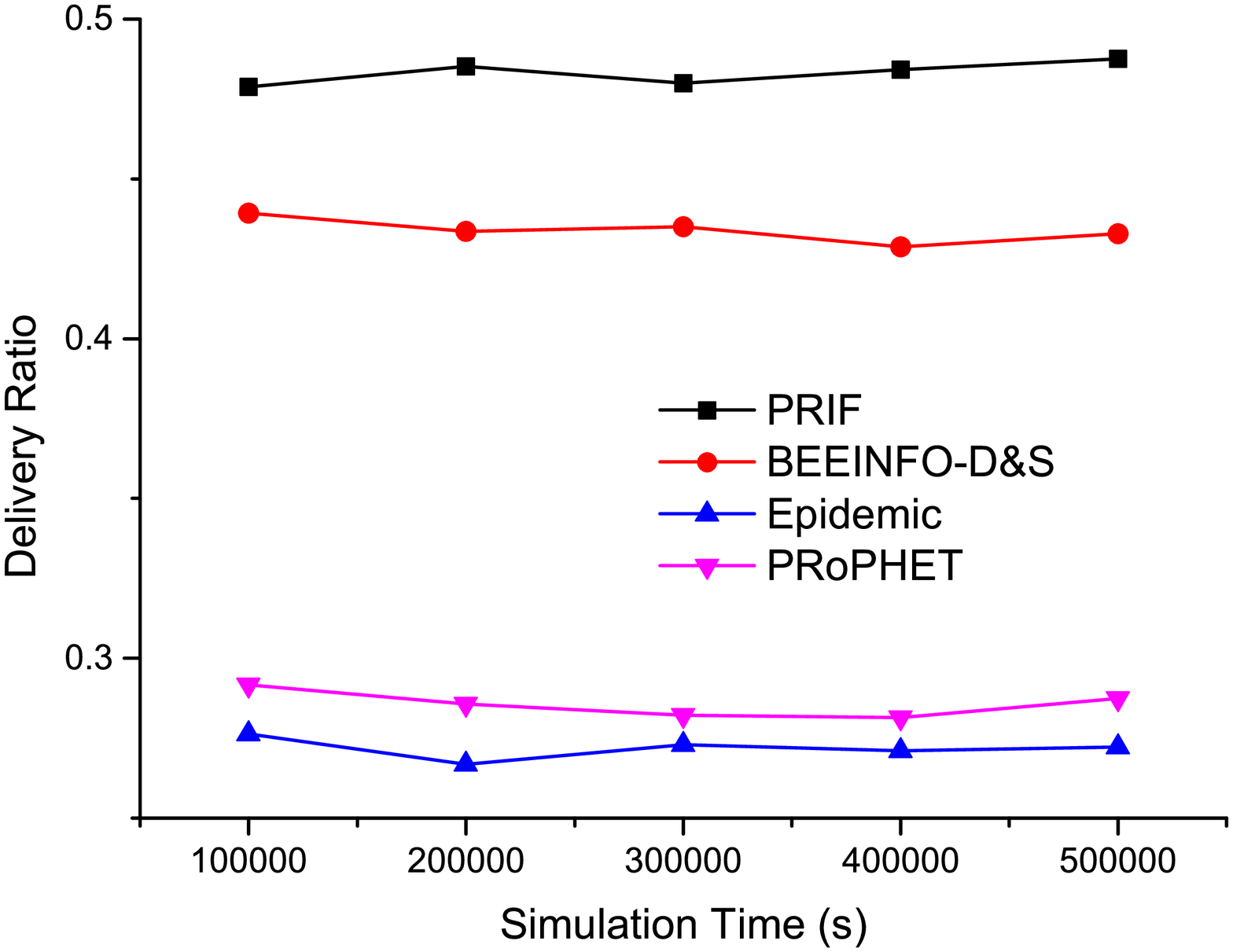}}
	~~
	\subfigure[]
	{\includegraphics[width=.30\textwidth]{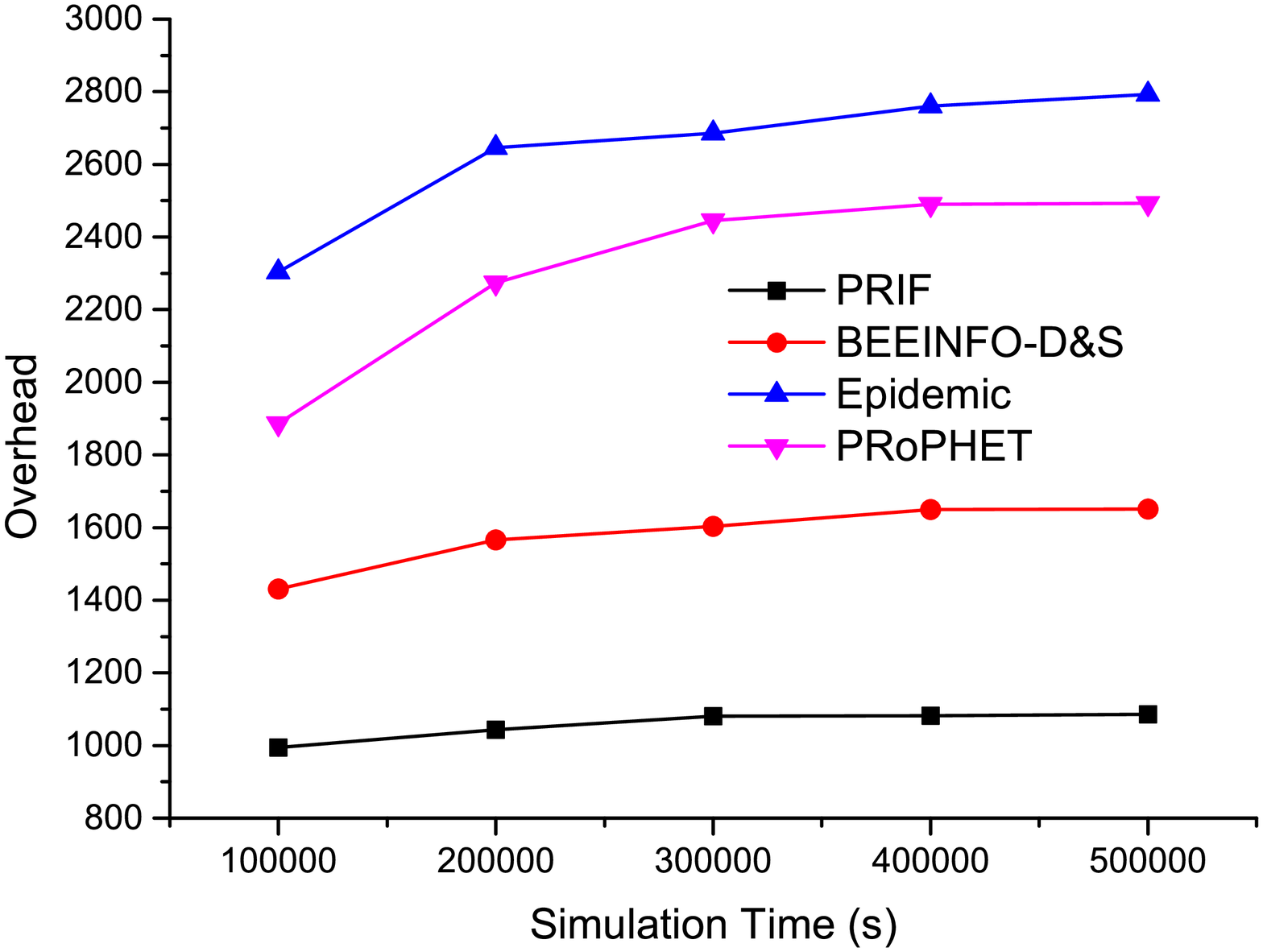}}
	~~
	\subfigure[]
	{\includegraphics[width=.30\textwidth]{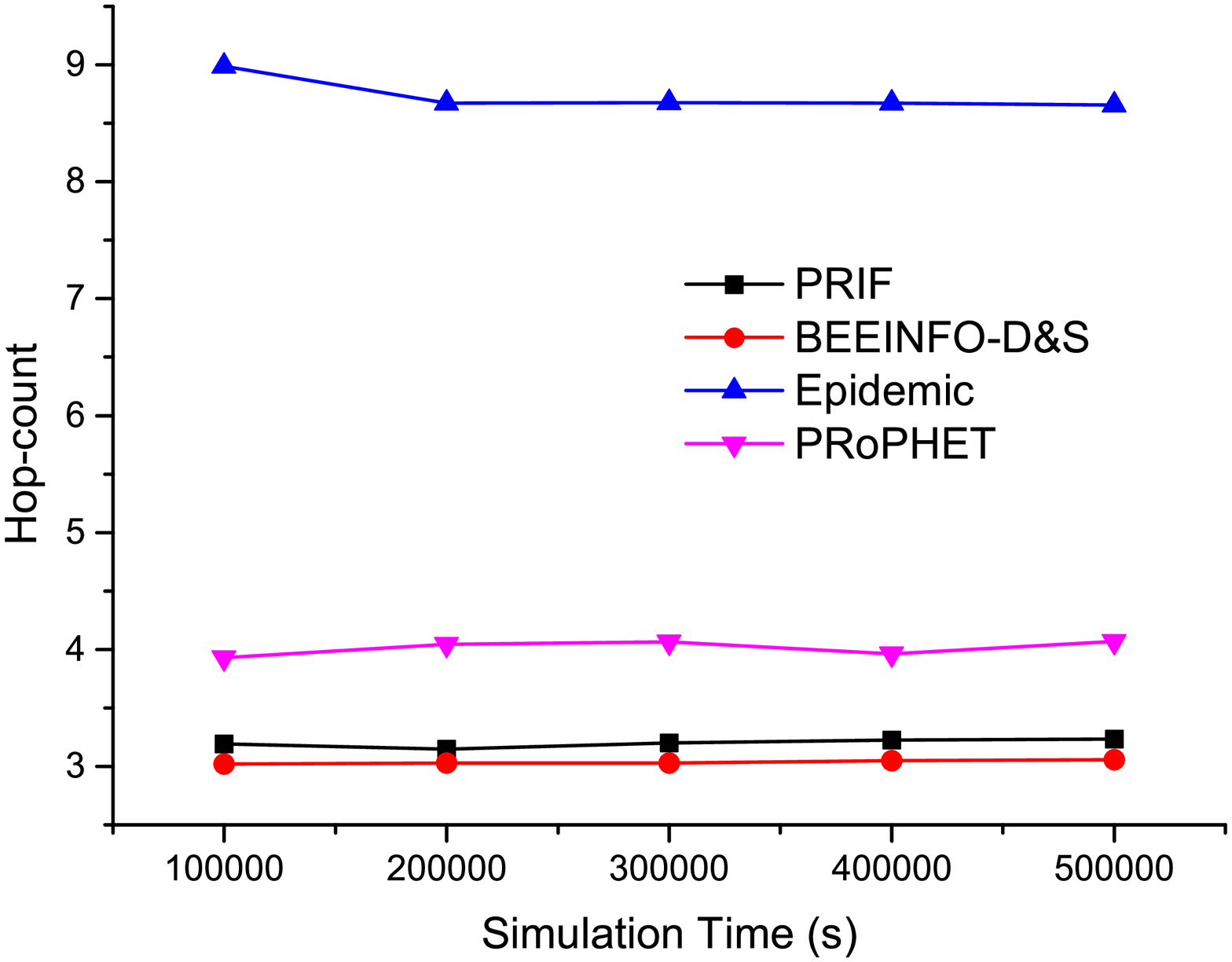}}
	~~
	\caption{Performance over simulation time.}
	\label{fig:Time}
\end{figure*}

The performance of the proposed PRIF is evaluated over different buffer sizes, message's TTL, and simulation time. Each experiment runs 30 times and we compute the average result. In Fig. \ref{fig:BS}, \ref{fig:TTL} and \ref{fig:Time}, we show the results of simulation experiments for delivery ratio, overhead, and hop count, respectively.

In Fig. \ref{fig:BS}, we compare the proposed PRIF scheme with other three schemes when buffer size ranges from 10 MB to 50 MB. It can be observed that, with larger buffer size, more messages will be delivered to the destinations, less overhead will be generated, and fewer hops are required. The proposed PRIF performs best in terms of the delivery ratio and overhead. For example, when the buffer size is set as 50 MB, PRIF delivers $65.29\%$ messages (compared with $62.57\%$ for BEEINFO-D$\&$S) with message overhead of 1146.5474 (compared with 1365.5567 for BEEINFO-$\&$S), and hop count of 2.8679 (similar to 2.7730 for BEEINFO-D$\&S$). By comparison, Epidemic and PRoPHET perform worse with $49.09\%$ and $43.84\%$ in delivery ratio, 1365.5567 and 2276.1788 in overhead, and 4.6457 and 3.2951 in hop count experiments respectively.

In Fig. \ref{fig:TTL}, we show the performance of the four schemes with varying TTL, where the simulation time is 400000 s and the buffer size is 10 MB. It can be seen that as the TTL increases, message delivery ratio of all schemes decreases, and PRIF exhibits best performance. When the TTL is set as 3600, PRIF delivers $42.75\%$ messages, which is $16.84\%$ higher than BEEINFO-D$\&$S, $75.42\%$ higher than Epidemic and $66.08\%$ higher than PRoPHET. For the overhead, PRIF also outperforms the rest. In terms of the hop count, the four schemes are comparable in performance, and PRIF is at similar level with that of BEEINFO-D$\&$S.

Fig. \ref{fig:Time} shows the performance of all these schemes varying with simulation time (using 10 MB of buffer size and 600 min of message TTL). When the simulation time increases from 100000 s to 500000 s, PRIF can gather more community energy information, which helps nodes to select better forwarders.
The trend of PRIF is similar to those of other schemes, but proves the over  all benefit as shown in Fig. \ref{fig:BS} and Fig. \ref{fig:TTL}. Hence, it can be concluded that, with long lifetime of networks, PRIF can give better performance while preserving the privacy of interests.

In summary, PRIF achieves better delivery ratio and overhead compared with the other three schemes, and gives comparable results with BEEINFO-D$\&$S for the hop count metric. These observations confirm the efficiency of introducing community energy in the design of social-based forwarding for SIOV.

\section{Conclusion}
In this paper, we propose a privacy-preserving interest-based forwarding scheme for SIOV, which not only protects nodes' interests, but also improves the forwarding performance. We have designed a privacy-preserving authentication protocol to recognize communities among mobile nodes. Moreover, we classify communities based on nodes' interests and present detailed methods to calculate community energy including inter-community energy and intra-community energy based on their interests. Extensive simulations have been conducted, which demonstrate the efficiency and effectiveness of the proposed scheme.

\ifCLASSOPTIONcaptionsoff
  \newpage
\fi

\bibliographystyle{ieeetr}

	\bibliography{mybibfile}

\begin{thebibliography}{10}

\bibitem{Yuan2013}
W.~Yuan, P.~Wang, W.~Liu, and W.~Cheng, ``Variable-width channel allocation for
  access points: {A} game-theoretic perspective,'' {\em {IEEE} Trans. Mob.
  Comput}, vol.~12, no.~7, pp.~1428--1442, 2013.

\bibitem{pelusi2006opportunistic}
L.~Pelusi, A.~Passarella, and M.~Conti, ``Opportunistic networking: data
  forwarding in disconnected mobile ad hoc networks,'' {\em {IEEE}
  Communications Magazine}, vol.~44, no.~11, pp.~134--141, 2006.

\bibitem{Lin2015}
X.~Lin and R.~Lu, {\em Vehicular Ad Hoc Network Security and Privacy}.
\newblock Wiley-IEEE Press, 2015.

\bibitem{Warthman2003}
F.~Warthman, {\em Delay-Tolerant Networks: A Tutorial}.
\newblock 2003.

\bibitem{vegni2015survey}
A.~M. Vegni and V.~Loscr{\`{\i}}, ``A survey on vehicular social networks,''
  {\em {IEEE} Communications Surveys and Tutorials}, vol.~17, no.~4,
  pp.~2397--2419, 2015.

\bibitem{Hu2015S}
X.~Hu, J.~Zhao, B.~Seet, V.~C.~M. Leung, T.~H.~S. Chu, and H.~C.~B. Chan,
  ``S-aframe: Agent-based multilayer framework with context-aware semantic
  service for vehicular social networks,'' {\em {IEEE} Trans. Emerging Topics
  Comput}, vol.~3, no.~1, pp.~44--63, 2015.

\bibitem{li2009localcom}
F.~Li and J.~Wu, ``Localcom: {A} community-based epidemic forwarding scheme in
  disruption-tolerant networks,'' in {\em Proceedings of the Sixth Annual
  {IEEE} Communications Society Conference on Sensor, Mesh and Ad Hoc
  Communications and Networks, {SECON} 2009, June 22-26, 2009, Rome, Italy},
  pp.~1--9, 2009.

\bibitem{musolesi2008writing}
M.~Musolesi, P.~Hui, C.~Mascolo, and J.~Crowcroft, ``Writing on the clean
  slate: Implementing a socially-aware protocol in haggle,'' in {\em 9th {IEEE}
  International Symposium on a World of Wireless, Mobile and Multimedia
  Networks, {WOWMOM} 2008, Newport Beach, CA, USA, 23-26 June, 2008}, pp.~1--6,
  2008.

\bibitem{musolesi2005adaptive}
M.~Musolesi, S.~Hailes, and C.~Mascolo, ``Adaptive routing for intermittently
  connected mobile ad hoc networks,'' in {\em 2005 International Conference on
  a World of Wireless, Mobile and Multimedia Networks {(WOWMOM} 2005), 13-16
  June 2005, Taormina, Italy, Proceedings}, pp.~183--189, 2005.

\bibitem{hui2007small}
P.~Hui and J.~Crowcroft, ``How small labels create big improvements,'' in {\em
  Fifth Annual {IEEE} International Conference on Pervasive Computing and
  Communications - Workshops (PerCom Workshops 2007), 19-23 March 2007, White
  Plains, New York, {USA}}, pp.~65--70, 2007.

\bibitem{hui2011bubble}
P.~Hui, J.~Crowcroft, and E.~Yoneki, ``{BUBBLE} rap: Social-based forwarding in
  delay-tolerant networks,'' {\em {IEEE} Trans. Mob. Comput}, vol.~10, no.~11,
  pp.~1576--1589, 2011.

\bibitem{daly2007social}
E.~M. Daly and M.~Haahr, ``Social network analysis for routing in disconnected
  delay-tolerant manets,'' in {\em Proceedings of the 8th {ACM} Interational
  Symposium on Mobile Ad Hoc Networking and Computing, MobiHoc 2007, Montreal,
  Quebec, Canada, September 9-14, 2007}, pp.~32--40, 2007.

\bibitem{bulut2010friendship}
E.~Bulut and B.~K. Szymanski, ``Friendship based routing in delay tolerant
  mobile social networks,'' in {\em Proceedings of the Global Communications
  Conference, 2010. {GLOBECOM} 2010, 6-10 December 2010, Miami, Florida,
  {USA}}, pp.~1--5, 2010.

\bibitem{li2014routing}
F.~Li, L.~Zhao, C.~Zhang, Z.~Gao, and Y.~Wang, ``Routing with multi-level
  cross-community social groups in mobile opportunistic networks,'' {\em
  Personal and Ubiquitous Computing}, vol.~18, no.~2, pp.~385--396, 2014.

\bibitem{xia2015beeinfo}
F.~Xia, L.~Liu, J.~Li, A.~M. Ahmed, L.~T. Yang, and J.~Ma, ``{BEEINFO:}
  interest-based forwarding using artificial bee colony for socially aware
  networking,'' {\em {IEEE} Trans. Vehicular Technology}, vol.~64, no.~3,
  pp.~1188--1200, 2015.

\bibitem{daly2009social}
E.~M. Daly and M.~Haahr, ``Social network analysis for information flow in
  disconnected delay-tolerant manets,'' {\em {IEEE} Trans. Mob. Comput},
  vol.~8, no.~5, pp.~606--621, 2009.

\bibitem{boldrini2008exploiting}
C.~Boldrini, M.~Conti, and A.~Passarella, ``Exploiting users' social relations
  to forward data in opportunistic networks: The hibop solution,'' {\em
  Pervasive and Mobile Computing}, vol.~4, no.~5, pp.~633--657, 2008.

\bibitem{epidemic}
A.~Vahdat and D.~Becker, ``Epidemic routing for partially-connected ad hoc
  networks,'' {\em Master Thesis}, 2000.

\bibitem{PRoPHET}
A.~Lindgren, A.~Doria, and O.~Schel{\'{e}}n, ``Probabilistic routing in
  intermittently connected networks,'' {\em Mobile Computing and Communications
  Review}, vol.~7, no.~3, pp.~19--20, 2003.

\bibitem{thing2006}
X.~You, Q.~Chen, B.~Fang, and Y.~Y. Tang, ``Thinning character using modulus
  minima of wavelet transform,'' {\em {International Journal of Pattern
  Recognition and Artificial Intelligence}}, vol.~20, no.~3, pp.~361--376,
  2006.

\bibitem{He2016Connected}
Z.~He, X.~Li, X.~You, D.~Tao, and Y.~Y. Tang, ``Connected component model for
  multi-object tracking,'' {\em {IEEE} Trans. Image Processing}, vol.~25,
  no.~8, pp.~3698--3711, 2016.

\bibitem{He2017Robust}
Z.~He, S.~Yi, Y.~Cheung, X.~You, and Y.~Y. Tang, ``Robust object tracking via
  key patch sparse representation,'' {\em {IEEE} Trans. Cybernetics}, vol.~47,
  no.~2, pp.~354--364, 2017.

\bibitem{Du2017A}
F.~Zhou, J.~Jin, X.~Du, B.~Zhang, and X.~Yin, ``A calculation method for social
  network user credibility,'' in {\em {IEEE} International Conference on
  Communications, {ICC} 2017, Paris, France, May 21-25, 2017}, pp.~1--6, 2017.

\bibitem{Du2016Social}
Y.~Zhang, F.~Tian, B.~Song, and X.~Du, ``Social vehicle swarms: a novel
  perspective on socially aware vehicular communication architecture,'' {\em
  {IEEE} Wireless Commun.}, vol.~23, no.~4, pp.~82--89, 2016.

\bibitem{Du2015improving}
P.~Dong, X.~Du, T.~Zheng, and H.~Zhang, ``Improving qos on high-speed vehicle
  by multipath transmission based on practical experiment,'' in {\em 2015
  {IEEE} Vehicular Networking Conference, {VNC} 2015, Kyoto, Japan, December
  16-18, 2015}, pp.~32--35, 2015.

\bibitem{Du2016Energy}
P.~Dong, X.~Du, J.~Sun, and H.~Zhang, ``Energy-efficient cluster management in
  heterogeneous vehicular networks,'' in {\em {IEEE} Conference on Computer
  Communications Workshops, {INFOCOM} Workshops 2016, San Francisco, CA, USA,
  April 10-14, 2016}, pp.~644--649, 2016.

\bibitem{DuC08}
X.~Du and H.~Chen, ``Security in wireless sensor networks,'' {\em {IEEE}
  Wireless Commun.}, vol.~15, no.~4, pp.~60--66, 2008.

\bibitem{du2009transactions}
X.~Du, M.~Guizani, Y.~Xiao, and H.-H. Chen, ``Transactions papers a
  routing-driven elliptic curve cryptography based key management scheme for
  heterogeneous sensor networks,'' {\em IEEE Transactions on Wireless
  Communications}, vol.~8, no.~3, pp.~1223--1229, 2009.

\bibitem{du2007effective}
X.~Du, Y.~Xiao, M.~Guizani, and H.-H. Chen, ``An effective key management
  scheme for heterogeneous sensor networks,'' {\em Ad Hoc Networks}, vol.~5,
  no.~1, pp.~24--34, 2007.

\bibitem{xiao2007survey}
Y.~Xiao, V.~K. Rayi, B.~Sun, X.~Du, F.~Hu, and M.~Galloway, ``A survey of key
  management schemes in wireless sensor networks,'' {\em Computer
  communications}, vol.~30, no.~11-12, pp.~2314--2341, 2007.

\bibitem{grant}
A.~C. B.~K. Vendramin, A.~Munaretto, M.~R. Delgado, and A.~C. Viana, ``Grant:
  Inferring best forwarders from complex networks' dynamics through a greedy
  ant colony optimization,'' {\em Computer Networks}, vol.~56, no.~3,
  pp.~997--1015, 2012.

\bibitem{BonehF01}
D.~Boneh and M.~K. Franklin, ``Identity-based encryption from the weil
  pairing,'' in {\em Advances in Cryptology - {CRYPTO} 2001, 21st Annual
  International Cryptology Conference, Santa Barbara, California, USA, August
  19-23, 2001, Proceedings}, pp.~213--229, 2001.

\bibitem{wu2014security}
L.~Wu, X.~Du, and X.~Fu, ``Security threats to mobile multimedia applications:
  Camera-based attacks on mobile phones,'' {\em IEEE Communications Magazine},
  vol.~52, no.~3, pp.~80--87, 2014.

\bibitem{wu2014mobifish}
L.~Wu, X.~Du, and J.~Wu, ``Mobifish: A lightweight anti-phishing scheme for
  mobile phones,'' in {\em Computer Communication and Networks (ICCCN), 2014
  23rd International Conference on}, pp.~1--8, IEEE, 2014.

\bibitem{huang2014achieving}
X.~Huang and X.~Du, ``Achieving big data privacy via hybrid cloud,'' in {\em
  Computer Communications Workshops (INFOCOM WKSHPS), 2014 IEEE Conference on},
  pp.~512--517, IEEE, 2014.

\bibitem{ONE-1}
Y.~Xian, C.~Huang, and J.~A. Cobb, ``Look-ahead routing and message scheduling
  in delay-tolerant networks,'' {\em Computer Communications}, vol.~34, no.~18,
  pp.~2184--2194, 2011.

\bibitem{ONE-2}
M.~Orlinski and N.~Filer, ``Neighbour discovery in opportunistic networks,''
  {\em Ad Hoc Networks}, vol.~25, pp.~383--392, 2015.

\bibitem{ONE-3}
J.~Miao, O.~Hasan, S.~B. Mokhtar, L.~Brunie, and G.~Gianini, ``A delay and cost
  balancing protocol for message routing in mobile delay tolerant networks,''
  {\em Ad Hoc Networks}, vol.~25, pp.~430--443, 2015.

\bibitem{ONE}
A.~Ker{\"{a}}nen, T.~K{\"{a}}rkk{\"{a}}inen, and J.~Ott, ``Simulating mobility
  and dtns with the {ONE} (invited paper),'' {\em {JCM}}, vol.~5, no.~2,
  pp.~92--105, 2010.

\end{thebibliography}

\begin{IEEEbiography}[{\includegraphics[width=1in,height=1.25in]{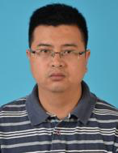}}]{Liehuang Zhu}
received his Ph.D. degree in computer science from Beijing Institute of Technology, Beijing, China, in 2004. He is currently a professor at the School of
Computer Science \& Technology, Beijing Institute of Technology. His research interests include security protocol analysis and design, group key exchange protocols, wireless sensor networks, and cloud computing.
\end{IEEEbiography}

\begin{IEEEbiography}[{\includegraphics[width=1in,height=1.25in,clip,keepaspectratio]{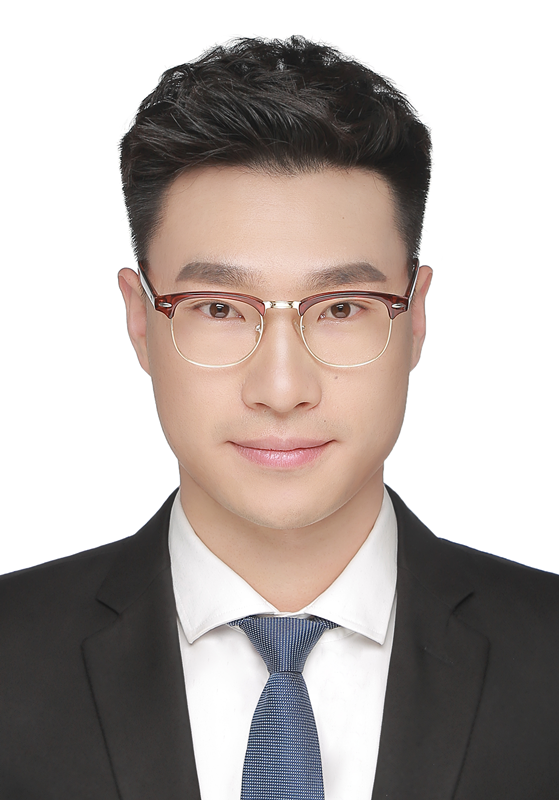}}]{Chuan Zhang}
received his bachelor's degree in network engineering from Dalian University of Technology, Dalian, China, in 2015. He is currently a Ph.D. student at the School of Computer Science and Technology, Beijing Institute of Technology. His current research interests include secure data services in cloud computing, security $\&$ privacy in VANET, and big data security.
\end{IEEEbiography}

\begin{IEEEbiography}[{\includegraphics[width=1in,height=1.25in,clip,keepaspectratio]{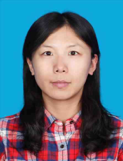}}]{Chang Xu}
	received her Ph.D. degree in computer science from Beihang University, Beijing, China, in 2013. She is currently an assistant professor at the School of Computer Science and Technology, Beijing Institute of Technology. Her research interests include security \& privacy in VANET, and big data security.
\end{IEEEbiography}

\begin{IEEEbiography}[{\includegraphics[width=1in,height=1.25in,clip,keepaspectratio]{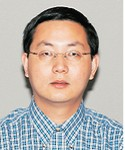}}]{Xiaojiang Du}
	is currently a professor in the Department of Computer and Information Sciences at Temple University. Dr. Du has published over 200 journals and conference papers in these areas and has been awarded more than \$5M research grants from the US National Science Foundation and Army Research Office. His research interests are security, systems, wireless networks, and computer networks.
\end{IEEEbiography}

\begin{IEEEbiography}[{\includegraphics[width=1in,height=1.25in,clip,keepaspectratio]{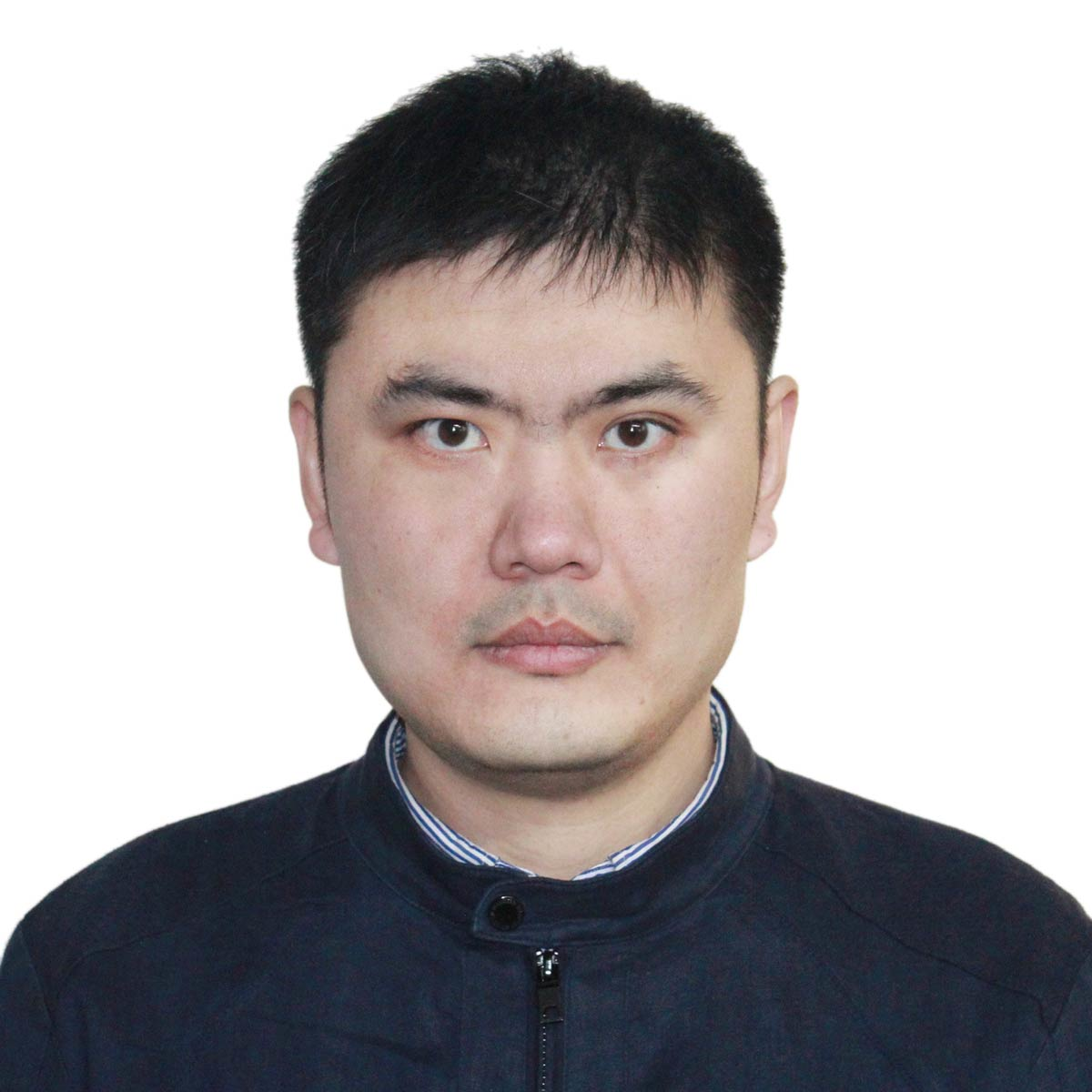}}]{Rixin Xu}
	received his bachelor's and master's degree in software engineering from Harbin Institute of Technology and Peking University respectively. He is currently pursuing his Ph.D. degree at the School of Computer Science and Technology, Beijing Institute of Technology. His current research interests include security of Internet of Things and Side-channel attacks.
\end{IEEEbiography}

\begin{IEEEbiography}[{\includegraphics[width=1in,height=1.25in,,clip,keepaspectratio]{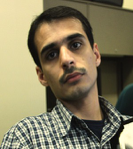}}]{Kashif Sharif}
	[M'$08$] received his MS degree in information technology in $2004$, and PhD degree in
	computing and informatics from University of North Carolina at Charlotte, USA in
	$2012$. He is currently an associate professor at Beijing Institute of
	Technology, China. His research interests include wireless \& sensor networks,
	network simulation systems, software defined \& data center networking, ICN,
	and Internet of Things. He is a member of IEEE and ACM.
\end{IEEEbiography}

\begin{IEEEbiography}[{\includegraphics[width=1in,height=1.25in,,clip,keepaspectratio]{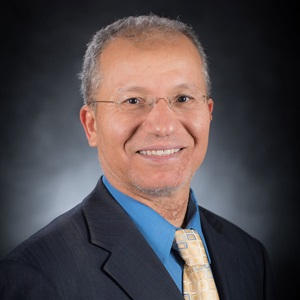}}]{Mohsen Guizani}
 has received his B.S., M.S., and Ph.D. from Syracuse University. He is currently a professor and the Electrical and Computer Engineering Department Chair at the University of Idaho. His research interests include wireless communications, mobile cloud computing, computer networks, security, and smart grid. He is the author of nine books and more than 400 publications. He was the Chair of the IEEE Communications Society Wireless Technical Committee. He served as an IEEE Computer Society Distinguished Speaker.
\end{IEEEbiography}

\end{document}